\documentclass[12pt]{article}
\usepackage{amsmath}
\usepackage{amssymb}
\begin{document}
\begin{center}
{\bf {\large{Reparameterization Invariant Model of a Supersymmetric System: BRST and Supervariable  Approaches}}}

\vskip 2.5cm

{\sf  A. Tripathi$^{(a)}$, B. Chauhan $^{(a)}$, A. K. Rao $^{(a)}$,  R. P. Malik$^{(a,b)}$}\\
$^{(a)}$ {\it Physics Department, Institute of Science,}\\
{\it Banaras Hindu University, Varanasi - 221 005, (U.P.), India}\\

\vskip 0.1cm

$^{(b)}$ {\it DST Centre for Interdisciplinary Mathematical Sciences,}\\
{\it Institute of Science, Banaras Hindu University, Varanasi - 221 005, India}\\
{\small {\sf {e-mails: ankur1793@gmail.com; bchauhan501@gmail.com;\\
amit.akrao@gmail.com;  rpmalik1995@gmail.com}}}
\end{center}

\vskip 1.5 cm

\noindent
{\bf Abstract:}
We carry out the Becchi-Rouet-Stora-Tyutin (BRST) quantization of the one (0 + 1)-dimensional (1D) model of a free massive spinning
relativistic particle (i.e. a supersymmetric system) by exploiting its {\it classical} infinitesimal and continuous 
reparameterization symmetry transformations. We use the
{\it modified} Bonora-Tonin (BT) supervariable approach (MBTSA) to BRST formalism to obtain the nilpotent (anti-)BRST symmetry
transformations of the target space variables and the (anti-)BRST invariant Curci-Ferrari 
(CF)-type restriction for the 1D  model of our
supersymmetric (SUSY) system. The nilpotent (anti-)BRST symmetry transformations for
 {\it other} variables of our model are derived by using
the (anti-)chiral supervariable approach (ACSA) to BRST formalism.
 Within the framework of the {\it latter}, we have shown the existence of the CF-type restriction by proving the (i)
 symmetry invariance of the coupled Lagrangians, and (ii) the absolute anticommutativity property of the 
 conserved (anti-)BRST charges. 
The application of the MBTSA to a {\it physical} SUSY system (i.e. a 1D model  of a massive spinning particle) is a 
{\it novel} result in our present endeavor. 
In the application of ACSA, we have considered {\it only} the (anti-)chiral super expansions of the
supervariables. Hence, the observation of the absolute anticommutativity of the (anti-)BRST charges is a novel result. 
The CF-type restriction is {\it universal} 
in nature as it turns out to be the {\it same} for the SUSY and non-SUSY reparameterization
 (i.e. 1D diffeomorphism) invariant models of the (non-)relativistic particles.



\newpage

\section {Introduction}

\noindent
The Becchi-Rouet-Stora-Tyutin (BRST) quantization scheme is one of the most elegant approaches to quantize the locally gauge {\it and} diffeomorphism invariant theories where the local {\it classical} transformation parameters are traded with the
(anti-)ghost fields at the {\it quantum} level [1-4]. For the quantization of the {\it classical} supersymmetric gauge theories
(with the bosonic and fermionic transformation parameters), the BRST quantization scheme requires the fermionic as well as the
bosonic (anti-)ghost fields/variables (see, e.g. [5, 6]). Some of the key characteristic features of the BRST quantization 
scheme are (i) for a  given {\it local} gauge and/or diffeomorphism symmetry, there exist {\it two} nilpotent symmetries
 which are christened as the BRST and anti-BRST
symmetries, (ii) the (anti-)BRST symmetries $(s_{(a)b})$ are fermionic (i.e. nilpotent) and absolutely 
anticommuting (i.e. $s_b\,s_{ab} + s_{ab}\,
s_b = 0$) in nature, (iii) the quantum gauge (i.e. BRST) invariance and unitarity are respected 
{\it together} in the perturbative computations  at any 
arbitrary order, (iv) there is appearance of the Curci-Ferrari (CF)-type restriction(s)
which are responsible for the absolute anticommutativity 
of the (anti-)BRST transformations {\it and} they lead to the existence of the coupled
(but equivalent) Lagrangian (densities) for the BRST {\it quantized}
theory which respect {\it both} (i.e. BRST and anti-BRST) nilpotent {\it quantum} symmetries, and (v) the (anti-)BRST
symmetries transform a bosonic field/variable to a fermionic field/variable and vice-versa. Hence, these symmetries are
 supersymmetric-type (i.e. SUSY-type).

Physically, the nilpotency property of the (anti-)BRST symmetry transformations corresponds to the 
{\it fermionic} nature of these {\it quantum}
symmetries {\it and} the absolute anticommutativity property encodes the linear independence of the BRST and anti-BRST symmetry 
transformations. The former property encodes the SUSY-type transformations. 
As pointed out earlier, the absolute anticommutativity property {\it owes} its dependence on the existence 
of the CF-type restriction(s). The BRST approach to Abelian 1-form gauge theory  is an {\it exception } where the CF-type restriction 
is {\it trivial} (but it turns out to be the limiting case of  the non-Abelian 1-form theory
 which is endowed with the CF-condition [7]).
It is the usual superfield approach (USFA) to BRST formalism [8-15] which provides the interpretation and origin for the abstract 
mathematical properties (i.e. nilpotency and anticommutativity) that are associated with the (anti-)BRST symmetries.
Furthermore, the USFA leads to the derivation of the CF-condition [7] in the context of a 4D 1-form non-Abelian  theory  
(see, e.g. [10-12]) which is found to be an (anti-)BRST invariant quantity. Hence, {\it it} is a {\it physical} restriction on the BRST {\it quantized} theory.

The USFA to BRST formalism [8-15] leads to the derivation of {\it only}  the off-shell nilpotent symmetries that 
are associated with the gauge and (anti-)ghost fields/variables. 
It does {\it not} shed {\it any} light on the (anti-)BRST transformations that are 
associated with the {\it matter} fields in an {\it interacting}
gauge theory. There have been consistent extensions of the USFA (see, e.g. [16-18]) where
 additional {\it quantum} gauge invariant restrictions 
on the superfields have been imposed to derive the (anti-)BRST symmetry 
transformations for the gauge, (anti-)ghost and {\it matter} fields 
{\it together}. This extended version of the superfield approach to BRST formalism has been called as the augmented version of
superfield approach (AVSA). In our recent works (see e.g. [19-21]), we have been able to 
develop a {\it simpler} off-shoot of AVSA where {\it only} the (anti-)chiral superfields/supervariables have been taken into account.
The {\it quantum} gauge [i.e. (anti-)BRST] invariant restrictions on these (anti-)chiral superfields/supervariables have led to the
deduction of (anti-)BRST symmetry transformations for {\it all} the fields/variables.
 This approach to BRST formalism has been named as the (anti-)chiral
superfield/supervariable approach (ACSA) to BRST formalism where the existence of the CF-type restriction(s) has been shown by proving
(i) the absolute anticommutativity of the (anti-)BRST charges, and (ii) the invariance 
of the coupled (but equivalent) Lagrangian (densities) of the (anti-)BRST invariant theories.

It has been a challenging problem to apply the superfield approach to BRST formalism in the context of (super)string and gravitational 
theories which are diffeomorphism invariant. In a recent paper [22], it has been proposed that
 the diffeomorphism symmetry can be taken into consideration within the framework of 
superfield approach to BRST formalism. This approach to BRST formalism has been called as the 
{\it modified} Bonora-Tonin superfield/supervariable approach (MBTSA) to BRST formalism which 
has been recently applied to the physical system of a 1D scalar
(non-)relativistic particles [23, 24]. To be precise, judicious combination of MBTSA and
 ACSA has been very fruitful in our recent works [23, 24] where 
we have been able to derive the proper (anti-)BRST symmetries for {\it all} the quantum
 variables along with the CF-type restriction in a systematic fashion. In 
the proposal by Bonora [26], the diffeomorphism symmetry transformations have 
been incorporated into the supefields which are defined on a (D, 2)-dimensional 
supermanifold on which a D-dimensional {\it ordinary} diffeomorphism invariant
 theory is generalized. We have exploited the mathematical rigor and beauty of the 
MBTSA in our present endeavor for the BRST analysis as well as quantization of 
a 1D diffeomorphism invariant SUSY system.

To be precise, in our present investigaion, we have applied the theoretical beauty and strength of MBTSA  
to derive the off-shell nilpotent (anti-)BRST transformations for the target space canonically conjugate  
{\it bosonic} variables ($x_\mu$ and $p^\mu$)  and {\it fermionic} variables $(\psi_\mu, \psi_5)$ along
with the (anti-)BRST invariant CF-type restrictions: $B + \bar B + i\,(\bar C\,\dot C - \dot {\bar C}\, C) = 0$
which is responsible for (i) the validity [cf. Eq. (22) below] of the absolute anticommutativity (i.e. $\{s_b,  s_{ab}\}
= s_b\, s_{ab} + s_{ab}\,s_b = 0$)  of the off-shell nilpotent ($s_{(a)b} ^2 = 0$) (anti-)BRST 
symmetry transformations $s_{(a)b}$, and (ii) the derivation of $L_B$ and $L_{\bar B}$ which are
coupled (but equivalent) [cf. Eq. (24) below]. The proper (anti-)BRST transformations of the {\it rest}
of the variables have been derived by using the ACSA to BRST formalism. It is worth pointing out that,
 in the case of MBTSA, we have taken into account the {\it full} super expansions of the 
supervariables along {\it all} the possible Grassmannian directions of the {\it general} 
(1, 2)-dimensional supermanifold. On the other hand, we have performed  {\it only} the (anti-)chiral super expansions of the 
supervariables in the context  of ACSA to BRST formalism. We have derived the {\it exact} expression for the
CF-type restriction by demanding (i) the invariance of the coupled (but equivalent) Lagrangians, and (ii) the validity 
 of the absolute anticommutatvity of the off-shell nilpotent (anti-)BRST symmetries 
as well as conserved (anti-)BRST charges in the {\it ordinary}
space {\it and} in the {\it superspace} (within the framework of ACSA to BRST formalism). 
These derivations of the CF-type restrictions,  by  various theoretical methods, are 
{\it novel} results in our present investigation.

The following key factors have been at the heart of our curiosity to pursue our present investigation.
First, we have been able to apply the MBTSA to a reparameterization invariant model of the {\it scalar}
relativistic particle to derive the (anti-)BRST symmetry transformations as well as the (anti-)BRST 
invariant CF-type of restriction [22]. Thus, it has been very important for us to apply the {\it same} mathematical 
technique (i.e. MBTSA) to a {\it physically} interesting SUSY model of a reparameterization
invariant theory where the fermionic as well as bosonic variables exist. Second, we have been very curious to
verify the {\it universality} of the CF-type restriction in the context of BRST quantization of the 1D diffeomorphism
(i.e. reparameterization) invariant theories. We find that the nature and form of the CF-type restriction is the {\it same} for the
SUSY as well as non-SUSY theories. Third, it has been very interesting to note that the gauge-fixing and Faddeev-Popov ghost terms {\it together} are same for the reparameterization invariant {\it scalar} and {\it SUSY} relativistic
as well as a non-relativistic  particle [23]. Finally, our present investigation (as well as others  [23, 24]) is our modest
initial steps to apply the MBTSA as well as the ACSA to BRST formalism {\it together} to physically interesting  4D (and 
higher dimensional) diffeomorphism invariant theories which are important from the point of view of the modern 
developments in gravitational as well as (super)string theories  (and related extended objects) of the
theoretical high energy physics.

The theoretical contents of our present endeavor are organised as follows. In Sec. 2, we discuss a couple of 
continuous and infinitesimal symmetry transformations and establish their relationship with the 
infinitesimal and continuous 1D diffeomorphism  (i.e. reparameterization) symmetry transformations.
Our Sec. 3 is devoted to the upgradation of the  {\it classical} infinitesimal reparameterization symmetry transformations 
to the {\it quantum} off-shell nilpotent and absolutely anticommuting (anti-)BRST symmetry transformations. The
{\it latter} property is satisfied due to the existence of an (anti-)BRST invariant CF-type restriction. This section 
 {\it also} contains the coupled (but equivalent) Lagrangians that respect {\it both} the (anti-)BRST symmetry transformations
on the submanifold (of the subspace of {\it quantum} variables) where the CF-type restriction is satisfied. In Sec. 4, we 
derive the (anti-)BRST transformations for the target space fermionic as well as bosonic variables. 
In addition, we deduce the CF-type of restriction  by exploiting the theoretical strength of MBTSA.
 We utilize the potential of ACSA to BRST 
formalism to obtain the nilpotent (anti-)BRST symmetry transformations for the {\it other} variables of our BRST-invariant
theory in Sec. 5. We capture the (anti-)BRST invariance of the coupled (but equivalent) Lagrangians within the ambit
of ACSA and establish, once again, the existence of our {\it quantum} (anti-)BRST invariant  CF-type restriction
in Sec. 6. Sec. 7 of our present endeavor  contains theoretical proof of the nilpotency and absolute anticommutativity of the conserved  and off-shell nilpotent (anti-)BRST charges. Finally, in Sec. 8, 
we summarize our key accomplishments and point out a few directions for further investigations.

In our Appendices A, B and C, we collect some explicit computations that corroborate a few key claims and statements 
that have been made in the main body of the text of our present endeavor. Our Appendix D is devoted to provide an alternative proof 
of the existence of an (anti-)BRST invariant CF-type restriction on our theory.\\

\noindent
{\it Convention and Notations}:  We take the flat metric tensor of the D-dimensional target spacetime manifold 
as $\eta_{\mu\nu} =  (+1, -1, -1, -1...)$ so that the dot product between {\it two} non-null vectors $(P_\mu, Q_\mu)$ is: 
$P\cdot Q = \eta_{\mu\nu}\,P^\mu\,Q^\nu = P_0\,Q_0 - P_i\,Q_i$ where the Greek indices $\mu, \nu, \lambda... = 0, 1, 2...D-1$
and Latin indices $i, j, k... = 1, 2, 3...D-1$. We adopt the convention of the left-derivative w.r.t. {\it all} the fermionic variables. 
We {\it always} denote the nilpotent (anti-)BRST transformations by the notations $s_{(a)b}$.
 Our 1D model is generalized onto a (1, 2)-dimensional supermanifold which is parameterized by the superspace
coordinates ${{Z}}^M = (\tau, \theta, \bar\theta)$ where $\tau$ is the bosonic evolution parameter and a pair of Grassmannian variable
$(\theta, \bar\theta)$ satisfy: $\theta^2 = {\bar\theta}^2 = 0, \; \theta\,\bar\theta + \bar\theta \, \theta = 0$. In our present
investigation, we shall focus {\it only} on the (1, 1)-dimensional (anti-)chiral super sub-manifolds 
of our chosen {\it general} (1, 2)-dimensional supermanifold in the context of ACSA.

\section {Preliminaries: Continuous and Infinitesimal Reparameterization Symmetry Transformations}

We divide our present section into two parts. We discuss, in our sub-section 2.1, a few {\it classical} infinitesimal 
continuous symmetries and their relationships with the {\it classical} infinitesimal reparameterization symmetry transformations. 
Our subsection 2.2 is devoted to a concise description of the {\it quantum} (anti-)BRST symmetry transformations.

\subsection{Some Classical Infinitesimal and Continuous symmetries}

We begin with the following {\it first-order} Lagrangian $(L_f)$ for the free one (0 + 1)-dimensional 
(1D) massive spinning (i.e. SUSY)  relativistic 
particle (see,  e.g. [5, 25])
\begin{eqnarray}
&&L_f = p_\mu\,\dot x^\mu  - \frac {e}{2}\;(p^2 - m^2) + \frac{i}{2}(\psi_\mu \,\dot\psi^\mu - \psi_5 \,\dot \psi_5)
+ i\,\chi \,(p_\mu \,\psi^\mu - m\,\psi_5), 
\end{eqnarray}
where we have parametrized the trajectory of the particle by $\tau$ and defined the ``generalized" velocities
$\big(\dot x^\mu = d\, x^\mu / {d\, \tau}, \;\dot \psi^\mu = d\, \psi^\mu /{d\, \tau}\big)$
 w.r.t. to {\it it}. This 1D trajectory is embedded in the D-dimensional flat Minkowskian target 
spacetime manifold where $(x_\mu, p^\mu)$ (with $\mu = 0, 1, 2,...D-1$) are the canonical conjugate 
pair of spacetime and momenta variables which are function of the evolution parameter $\tau$.
We have the fermionic variables $(\chi, \psi_\mu, \psi_5)$ in our theory. The Lagrangian (1) {\it also} has the bosonic variable  $e$ and fermionic variable $\chi$ as the Lagrange multiplier variables. These {\it latter} variables behave like the ``gauge" and ``superaguge" variables due to their transformation properties under the 
gauge and supersymmetric gauge transformations. 
In fact, the fermionic variable $ \psi_\mu$ is the superpartner of $x_\mu$ and the other fermionic
variable $ \psi_5 $ has been introduced to incorporate the rest mass $m$ into the Lagrangian $ L_f $
where the mass-shell condition $ (p^2\,-\,m^2=0)$ is satisfied by the free ($ \dot p_{\mu} = 0 $) massive
spinning relativistic particle. Our present 1D model is interesting because it provides a prototype example
of a supersymmetric (SUSY) gauge theory and its generalization to the 4D theory provides an example of the 
supergravity theory where fermionic variable $\psi_{\mu}$ becomes the Rarita-Schwinger Lorentz vector spin 3/2 field and the einbein variable $e$ turns {\it itself} into the vierbein field.

The action integral $S = \int_{- \infty}^{+ \infty} d\,\tau \,L_f$ respects the following 
infinitesimal and continuous reparameterization symmetry transformations $(\delta_r)$
\begin{eqnarray}
&&\delta_r \,x_\mu  = \epsilon\, {\dot x}_\mu, \;\;\; \qquad \delta_r \psi_\mu = \epsilon\, {\dot\psi}_\mu, 
\qquad \qquad \quad \delta_r\, p_\mu = \epsilon\,{\dot p}_\mu,
 \nonumber \\
&&\delta_r\, \psi_5 = \epsilon\, {\dot\psi}_5, \qquad \;\;\;  
\;\;\;\delta_r\, \chi = \frac{d}{d\,\tau}\,(\epsilon\,\chi),   \qquad \quad
 \delta_r\, e  = \frac{d}{d\,\tau}\,(\epsilon\,e),
\end{eqnarray}
because the first-order Lagrangian $(L_f)$ transforms, under the above infinitesimal
reparameterization symmetry transformation $(\delta_r)$, as
\begin{eqnarray}
\delta_r \, L_f = \frac{d}{d\tau} \, [\epsilon\,L_f]\,\,\quad \Longrightarrow \,\,\quad \delta_r\,S = 0,
\end{eqnarray}
where $\delta_r$ {\it basically} corresponds to the infinitesimal 1D diffeomorphism/reparameterization transformation: 
$ \tau \longrightarrow \tau' = \tau\,-\,\epsilon(\tau)$. Here the transformation parameter $\epsilon(\tau)$ is infinitesimal.
It is an elementary exercise to note that, if we set all the fermionic variables equal to zero (i.e. $\chi, \psi_\mu, \psi_5 = 0$), we 
obtain an infinitesimal gauge symmetry transformation $(\delta_g)$ from the infinitesimal reparameterization symmetry transformation
(2) as follows 
\begin{eqnarray}
\delta_g x_\mu  = \xi\, p_\mu, \;\;\;\quad \delta_g p_\mu = 0,\;\;\;\quad \delta_g e  = \frac{d}{d\,\tau}(\xi) = \dot \xi,\;\;\;
\quad \delta_g \psi_\mu = \delta_g \psi_5 = \delta_g \chi = 0,  
\end{eqnarray}
where we have identified $e\,\epsilon = \xi$ and used the Eular-Lagrange eqations of motions (EL-EOMs):
${\dot x}_\mu = e\,p_\mu, \;{\dot p}_\mu = 0$. In equation (4),  the bosonic
infinitesimal transformation parameter $\xi(\tau)$ is nothing but the gauge transformation parameter. 
It can be readily checked that we have the following transformation for $L_f$ and S under $\delta_g$:
\begin{eqnarray}
\delta_g \, L_f = \frac{d}{d\tau} \, \Big[\frac{\xi}{2}\,(p^2 + m^2)\Big] \quad \implies \quad \delta_g \, 
S = 0, \qquad S =  \int_{-\infty}^{+\infty}d\tau \;L_f. 
\end{eqnarray}
We have an infinitesimal classical supergauge symmetry transformations $(\delta_{sg})$ in our theory which transforms
the {\it fermionic} variables into their {\it bosonic} counterparts and vice-versa. These continuous and infinitesimal symmetry transformations are
\begin{eqnarray}
&&\delta_{sg} \, x_\mu = \kappa \, \psi_\mu, \qquad \qquad \delta_{sg} \, p_\mu = 0, \qquad \qquad \delta_{sg} \, \psi_\mu = 
i\,\kappa \, p_\mu, \nonumber \\ 
&&\delta_{sg} \,\psi_5 = i\,\kappa \,m, \qquad \qquad  \delta_{sg} \, \chi = 
i\,\dot\kappa, \qquad \qquad \delta_{sg} \, e = 2\,\kappa \, \chi,
\end{eqnarray}
where the infinitesimal transformation parameter $\kappa(\tau)$ is {\it fermionic} (i.e. $\kappa^2 = 0$) in nature.
It is straightforward to observe that we have the following:  
\begin{eqnarray}
\delta_{sg} \, L_f = \frac{d}{d\tau} \, \Big[\frac{\kappa}{2}\,
(p_\mu \, \psi^\mu + m \, \psi_5)\Big]\quad \implies \quad \delta_{sg} \, S = 0. 
\end{eqnarray}
Under the combined $\delta = \delta_g + \delta_{sg}$ {\it classical} symmetry transformation $(\delta)$, 
we have the following continuous and infinitesimal symmetry transformations $(\delta)$, namely; 
\begin{eqnarray}
&&\delta\,x_\mu = \xi \, p_\mu +\kappa\, \psi_\mu, \qquad\delta\psi_\mu = i\,\kappa\,p_\mu, \qquad \delta p_\mu = 0,\nonumber\\
&&\delta e = \dot\xi +2\,\kappa\,\chi,\qquad\delta\chi  = i\,\dot\kappa,\qquad \delta\psi_5 = i\,\kappa\,m,
\end{eqnarray}
which lead to the transformation of the first-order Lagrangian $L_f$ as
\begin{eqnarray}
\delta L_f  = \frac {d}{d\tau}\,\Big[\frac{\xi}{2}\,(p^2 + m^2) + \frac{\kappa}{2}\,(p_\mu\,\psi^\mu 
+ m\, \psi_5)\Big] \quad \implies \quad \delta \, S = 0.
\end{eqnarray}
Thus, the continuous and infinitesimal transformation $\delta$ is indeed a {\it symmetry} transformation for the action integral
$S = \int_{-\infty}^\infty  d\,\tau\,L_f$ due to Gauss's divergence theorem.

As the gauge symmetry transformation (4) can be incorporated into the reparameterization 
symmetry transformations (2) with the help of some EL-EOMs and some identification of the
transformation parameters, in exactly similar fashion, the combined (super)gauge symmetry transformations
(8) can be incorporated into the reparameterization symmetry transformations (2) if we take the help
of the following EL-EOMs, namely;
\begin{eqnarray}
\dot p_\mu = 0, \qquad \dot x_\mu = e\,p_\mu - i\,\chi\,\psi_\mu,\qquad \dot \psi_\mu = \chi\, p_\mu,\qquad \dot \psi_5 = m\,\chi,
\end{eqnarray}
and identify the transformation parameters as: $ e\,\epsilon = \xi$ and $-\,i\, \epsilon\,\chi = \kappa$ (see, e.g. [5, 25] for details). 
Thus, we note that the {\it classical} infinitesimal reparameterization symmetry transformations (2) are a set of very
{\it general} kind of symmetry transformations whose {\it special} cases are the continuous and infinitesimal symmetry transformations
(4) and (8).

\subsection{Quantum (Anti-)BRST Symmetries Corresponding to the Classical (Super)gauge Symmetry Transformations}

The {\it classical} continuous and infinitesimal (super)gauge symmetry transformations (8) can be elevated to their counterpart
{\it quantum} nilpotent $(s_{(a)b}^2 = 0)$, infinitesimal and continuous (anti-)BRST transformations 
$s_{(a)b}$ as follows [5, 6, 25]
\begin{eqnarray} 
&& s_{ab}\; x_\mu = {\bar c}\; p_\mu + \bar \beta \;\psi_\mu,\;\quad\qquad s_{ab} \;\psi_\mu = i \;\bar \beta\; p_\mu, \quad\qquad s_{ab}\; e = \dot {\bar c} + 2 \;\bar \beta\; \chi,  
\nonumber\\
&& s_{ab}\; c = i\; \bar b,\quad s_{ab}\; \bar c = - i \;{\bar \beta}^2, \quad s_{ab}\; p_\mu = 0 \quad s_{ab}\; \bar \beta = 0, 
\;\quad s_{ab} \; \beta = - i\; \gamma, \nonumber\\
&& s_{ab} \;\gamma = 0, \quad s_{ab}\;\chi = i\; \dot {\bar \beta}, 
\quad s_{ab} \; b =  2\; i\; \bar \beta\; \gamma,\quad s_{ab} \,\psi _5 = i\,\bar\beta\,m, \quad s_{ab}\; \bar b = 0,
\end{eqnarray}
\begin{eqnarray}
&&s_b\; x_\mu = c\;p_\mu + \beta \;\psi_\mu, \quad\qquad s_b\; \psi_\mu = i\;\beta\; p_\mu, \quad\qquad s_b\; e = \dot c + 2\;\beta\; \chi,  
\nonumber\\
&& s_b\; p_\mu = 0,\quad s_b\;c = - i\; \beta^2,  \;\quad s_b \;\beta = 0,\;\quad s_b \;{\bar c} = i\; b, 
\;\quad s_b \;\bar \beta = i \;\gamma,\nonumber\\
&& s_b \;\gamma = 0,  \quad s_b \;\chi = i\; \dot \beta, 
\qquad s_b\; \bar b = - 2\; i\; \beta\; \gamma,\quad s_b \;b = 0, \quad s_{b}\, \psi _5 = i\,\beta\,m,
\end{eqnarray} 
where the {\it fermionic} $(c^2 = {\bar c}^2 = c\,{\bar c} + {\bar c}\,c = 0)$ (anti-)ghost variables $(\bar c)c$ and the
{\it bosonic} $(\beta^2 = {\bar\beta}^2 \neq 0, \;\beta\,{\bar\beta} =  {\bar\beta}\,\beta)$ (anti-)ghost variables $(\bar\beta)\beta$
correspond to the bosonic and fermionic gauge and supergauge transformation parameters $\xi$ and $\kappa$ of Eq. (8), respectively. 
The variables $(\bar b)b$ are the Nakanishi-Lautrup type auxiliary variables and $\gamma$ is an additional fermionic 
$(\gamma^2 = 0)$ variable in our BRST-quantized  (as well as invariant) theory.

It is straightforward to note that the anticommutativity property of the off-shell
nilpotent $(s_{(a)b}^2 = 0)$ (anti-)BRST symmetry transformations ($s_{(a)b}$), namely;
\begin{eqnarray}
\{s_b, s_{ab}\}\,x_\mu = i\,(b + \bar b + 2\,\beta\,\bar\beta)\,p_\mu, \qquad \quad \{s_b, s_{ab}\}\,e = i\,\frac{d}{d\,\tau}\,
(b + \bar b + 2\,\beta\,\bar\beta), 
\end{eqnarray}
is {\it true} if and only if the CF-type restriction: $b + \bar b + 2\,\beta\,\bar\beta = 0$ is invoked from {\it outside}. However, this
restriction is a {\it physical} constraint because {\it it} is an (anti-)BRST invariant (i.e. $s_{(a)b}\,
[b + \bar b + 2\,\beta\,\bar\beta] = 0$) quantity. It can be readily checked that:
\begin{eqnarray}
\{s_b, s_{ab}\}\,\Phi = 0, \qquad\quad \Phi = p_\mu, \psi_\mu, \psi_5, b, \bar b, \beta, \bar\beta, c, \bar c, \gamma.   
\end{eqnarray}
In other words, we observe that the off-shell nilpotent $(s_{(a)b}^2)  = 0$ (anti-)BRST symmetry transformations $(s_{(a)b})$ are absolutely
anticommuting (i.e. $\{s_b, s_{ab}\} = s_b\,s_{ab} + s_{ab}\,s_b = 0$) provided the entire theory is considered on the quantum submanifold 
where the CF-type restriction $b + \bar b + 2\,\beta\,\bar\beta = 0$ is satisfied. It is the existence of {\it this} physical 
restriction that leads to the existence of coupled (but equivalent) Lagrangians
\begin{eqnarray}
&&L_b = L_f + s_b\,s_{ab}\,\Big[\frac{i}{2}\,e^2 + \chi\,\psi_5 -\frac {1}{2} \bar c\,c \Big], \nonumber\\
&&L_{\bar b} = L_f - s_{ab}\,s_{b}\,\Big[\frac{i}{2}\,e^2 + \chi\,\psi_5 -\frac{1}{2} \bar c\,c \Big],
\end{eqnarray}
which incorporate the gauge-fixing and Faddeev-Popov ghost terms in addition to the first-order Lagrangian
$(L_f)$ of Eq. (1). In the full blaze of glory, the above Lagrangians (in terms of {\it all} the appropriate variables) are as follows [6, 25]
\begin{eqnarray}
 L_b & = & L_f + b\,({\dot e} + 2\, {\bar \beta}\,\beta)  + b^2 - i\,\dot{\bar c} \, {\dot c} + {\bar \beta}^2 \, {\beta}^2 - 2\,e\,(\bar\beta \, \dot\beta + \gamma\,\chi)
+ 2\, i\, \chi\,(\beta\,\dot{\bar c} - \bar{\beta} \, \dot c)   \nonumber\\
& + & m \, (\bar\beta \, \dot\beta - \dot{\bar \beta}\, \beta + \gamma \, \chi) +  2\,\gamma\,(\beta \, \bar c - \bar\beta \, c) 
- \dot \gamma \, \psi_5, 
\end{eqnarray}
\begin{eqnarray}
L_{\bar b} & = & L_f  - \bar b\,({\dot e} - 2\, {\bar \beta}\,\beta) + {\bar b}^2 - i\,\dot{\bar c} \, {\dot c} + {\bar \beta}^2 \, {\beta}^2 
+ 2\, i\, \chi\,(\beta\,\dot{\bar c} - \bar{\beta} \, \dot c)  + 2\,e\,(\dot{\bar \beta} \, \beta - \gamma\,\chi) \nonumber\\
&+& m \, (\bar\beta \, \dot\beta - \dot{\bar \beta}\, \beta + \gamma \, \chi) +  2\,\gamma\,(\beta \, \bar c - \bar\beta \, c) 
- \dot \gamma \, \psi_5,
\end{eqnarray}
where the subscripts $b$ and $\bar b$ are appropriate because the Lagrangian $L_b$ depends on the Nakanishi-Lautrup 
auxiliary variable $b$ {\it but} the Lagrangian $L_{\bar b}$ contains the auxiliary variable $\bar b$ in its full expression. It is straightforward
to check that $L_b$ and $L_{\bar b}$ of our theory respect the {\it perfect} BRST and anti-BRST transformations because we have:
\begin{eqnarray}
s_b \, L_b = \frac{d}{d\,\tau}\,\Big[ \frac{\beta}{2}\,(p_\mu \, \psi^\mu + m \, \psi_5) + \frac{c}{2}\,(p^2 + m^2) 
+ b \, (\dot c + 2\,\beta\, \chi)\Big],   
\end{eqnarray}
\begin{eqnarray}
s_{ab} \, L_{\bar b} = \frac{d}{d\,\tau}\,\Big[\frac{\bar\beta}{2}\,(p_\mu \, \psi^\mu + m \, \psi_5) + \frac{\bar c}{2}\,(p^2 + m^2)  
- \bar b \, (\dot{\bar c} + 2\,\bar\beta \, \chi)\Big].
\end{eqnarray}
As a consequence, the action integrals $S_1 = \int^\infty_{- \infty} d\,\tau\, L_b$ and 
$S_2 = \int^\infty_{- \infty} d\,\tau \,L_{\bar b}$ remain invariant under the BRST and anti-BRST
symmetry transformations (12) and (11), respectively. We define a {\it perfect} symmetry as the one under 
which the action integral remains  invariant {\it without} any use of the CF-type restriction and/or EL-EOMs.

The BRST quantization of the massive spinning particle can be performed using the (anti-)BRST transformations 
(11) and (12) which correspond to the {\it classical} (super)gauge symmetry transformations (8). In our recent
publication [25], we have discussed {\it all} the details of this quantization scheme. However, 
we have {\it not} touched the continuous
and infinitesimal reparameterization transformations (2). We focus on the {\it latter} {\it classical} symmetry transformations
in the {\it next} section for the BRST analysis as it is our modest {\it first} step towards 
our main goal to discuss the diffeomorphism invariant SUSY 
theories in the physical (3 + 1)-dimensional (4D) and higher dimensional $(D > 4)$ spacetime.

\section{Quantum (Anti-)BRST Symmetries Corresponding to the Classical Reparameterization Symmetry}

In this section, we discuss the {\it quantum} (anti-)BRST symmetry transformations corresponding to the {\it classical} infinitesimal
reparameterization symmetry transformations (2). This is essential and important because we wish to perform the BRST quantization of a
1D diffeomorphism (i.e. reparameterization) invariant SUSY theory. We exploit the standard techniques and tricks of the 
BRST formalism to elevate the above {\it classical} symmetry to its counterparts {\it quantum} symmetries. In fact,   
the off-shell nilpotent $(s_{(a)b}^2 = 0)$ (anti-)BRST symmetry
transformations [corresponding to the {\it classical} Eq. (2)] are 
\begin{eqnarray} 
&& s_{ab} \;\psi_\mu = {\bar C}\; {\dot\psi}_\mu,  \qquad s_{ab}\; p_\mu = {\bar C}\; {\dot p}_\mu, \qquad
s_{ab}\; e = \frac{d}{d\,\tau}\,({\bar C}\,e), \quad s_{ab}\; x_\mu = {\bar C}\; {\dot x}_\mu, \nonumber\\  
&&s_{ab}\; \bar C = \bar C \, \dot{\bar C}, \qquad 
s_{ab}\;\chi = \frac{d}{d\,\tau}\,({\bar C}\,\chi),\quad s_{ab} \,\psi _5 = {\bar C}\; {\dot\psi}_5,  \qquad 
s_{ab}\; C = i\,\bar B, \nonumber\\
&& s_{ab}\; \bar B = 0, \qquad\qquad s_{ab} \; B =  {\dot B}\,{\bar C} - B\,\dot{\bar C},
\end{eqnarray}
\begin{eqnarray}
&& s_{b}\; x_\mu = {C}\; {\dot x}_\mu, \qquad s_{b}\; p_\mu = {C}\; {\dot p}_\mu, \qquad
s_{b}\; e = \frac{d}{d\,\tau}\,({C}\,e), \qquad s_{b} \;\psi_\mu = {C}\; {\dot\psi}_\mu, \nonumber\\  
&&s_{b} \,\psi _5 = {C}\; {\dot\psi}_5, \qquad 
s_{b}\;\chi = \frac{d}{d\,\tau}\,({C}\,\chi), \qquad s_{b}\; \bar C = i\,B, \;\;\; \qquad 
 s_{b}\;  B = 0,\nonumber\\
&&s_{b}\; C = C\,{\dot C},  \qquad\qquad s_{b} \; \bar B =  \dot{\bar B}\,{C} - \bar B\,\dot{C},
\end{eqnarray} 
where $B$ and $\bar B$ are the Nakanishi-Lautrup auxiliary and $({\bar C})C$ are the 
(anti-)ghost variables of our theory. As far as the absolute
anticommutativity property (i.e. $\{s_b, s_{ab}\} = 0$) of the above transformations
 is concerned, we note the following
\begin{eqnarray}
&&\{s_b, s_{ab}\}\,s_\mu = i\,\big[B + {\bar B} + i\,({\bar C}\,{\dot C} - \dot{\bar C}\,C)\big]\,{\dot s}_\mu, \nonumber\\
&&\{s_b, s_{ab}\}\,\Phi = i\,\frac{d}{d\,\tau}\Big[\big\{B + {\bar B} + i\,({\bar C}\,{\dot C} - \dot{\bar C}\,C\big\}\,\Phi\Big], \nonumber\\
&&\{s_b, s_{ab}\}\,\Psi = 0, \qquad \Psi = B, \bar B, C, \bar C,
\end{eqnarray}
where $ s_{\mu}= x_\mu (\tau),\,p_\mu (\tau),\,\psi_\mu (\tau),\,\psi_5 (\tau)$ and $\Phi= e (\tau),\, \chi (\tau)$. Thus, we note that
the absolute anticommutativity property of the (anti-)BRST symmetry transformations [cf. Eqs. (20),(21)]
is satisfied (i.e. $\{s_b, s_{ab}\}= s_b\,s_{ab}\,+\,s_{ab}\,s_b = 0$ ) if and only if we {\it invoke} 
the (anti-)BRST invariant (i.e. $s_{(a)b}\,[B + {\bar B} + i\,({\bar C}\,{\dot C} - \dot{\bar C}\,C)]=0$)
CF-type restriction  $[B + {\bar B} + i\,({\bar C}\,{\dot C} - \dot{\bar C}\,C) = 0]$. We note, therefore, that a 
CF-type constraint exists on our theory which is the root-cause behind the absolute anticommutativity 
(i.e. $\{s_b, s_{ab}\}=0$) of the (anti-)BRST symmetry transformations {\it and} it leads to  the existence of the 
coupled (but equivalent) (anti-)BRST invariant Lagrangians as
\begin{eqnarray}+
&&L_B = L_f + s_b\,s_{ab}\,\Big[\frac{i}{2}\,e^2 + \chi\,\psi_5 - \frac {1}{2}\,\bar C\,C \Big], \nonumber\\
&&L_{\bar B} = L_f - s_{ab}\,s_{b}\,\Big[\frac{i}{2}\,e^2 + \chi\,\psi_5 - \frac {1}{2}\,\bar C\,C \Big],
\end{eqnarray}
where the (anti-)BRST symmetry transformations $ (s_{(a)b})$ are the {\it quantum} symmetries [cf. Eqs. (20), (21)]  
corresponding to the {\it classical} infinitesimal reparameterization symmetry transformations
(2). It will be noted that the quantities in the square brackets of (23) are the {\it same} as
quoted in Eq. (15). However, the (anti-)BRST symmetry transformations in (23) are {\it different}
from (15) as {\it are} the notations for the Nakanishi-Lautrup type auxiliary and (anti-)ghost variables [cf.  Eqs. (11), (12),
(20) and (21) for details].

The above coupled (but equivalent) Lagrangians (23) can be expressed in terms of the {\it auxiliary} and {\it basic} variables
in an explicit form as:
\begin{eqnarray}
&&L_B = L_f  + B\,\Big[e\,\dot e -i\, (2\, \dot{\bar C}\, C  + {\bar C}\,\dot C)\Big]+ \frac{B^2}{2} -\,i\,e^2\,\dot{\bar C}\,\dot C 
- i\,e\,\dot e\,\dot{\bar C}\,C - \dot{\bar C}\,{\bar C}\,\dot C\,C, \nonumber\\
&&L_{\bar B} = L_f  - \bar B\,\Big[e\,\dot e - i\,(2\, {\bar C}\,\dot C +  \dot{\bar C}\,C)\Big]
+\frac{{\bar B}^2}{2}  -\,i\,e^2\,\dot{\bar C}\,\dot C - i\,e\,\dot e\,{\bar C}\,\dot C - \dot{\bar C}\,{\bar C}\,\dot C\,C.
\end{eqnarray}
We note that the pure Faddeev-Popov ghost part (i.e. $-\, \dot{\bar C}\,{\bar C}\,\dot C\,C$) of the above coupled (but equivalent)
Lagrangians is the {\it same}. It can be readily checked 
that the EL-EOMs from Lagrangians $L_B$ and $L_{\bar B}$, w.r.t. the auxiliary variables $B$ and $\bar B$, lead to the following
\begin{eqnarray}
B = -\,e\,\dot e +2\,i\,\dot{\bar C}\,C + i\,\bar C\,\dot C, \qquad \bar B = e\,\dot e - 2\,i\,{\bar C}\,\dot C - i\,\dot{\bar C}\, C, 
\end{eqnarray}
which are responsible for the derivation of the CF-type restriction: $B + \bar B + i\,(\bar C\,\dot C - \dot{\bar C}\,C) = 0$. The above Lagrangians
$L_B$ and $L_{\bar B}$ of our theory respect the BRST and anti-BRST transformations in a precise and {\it perfect} manner because it is interesting 
to check that:
\begin{eqnarray}
&&s_b\,L_B = \frac{d}{d\,\tau}\Big[C\,L_f + e^2\,B\,\dot C + e\,\dot e\,B\,C - i\,B\,\bar C\,\dot C\,C + B^2\,C\Big],
\end{eqnarray}
\begin{eqnarray}
&&s_{ab}\,L_{\bar B} = \frac{d}{d\,\tau}\Big[\bar C\,L_f - e^2\,\bar B\,\dot{\bar C} - e\,\dot e\,\bar B\,\bar C 
- i\,\bar B\,\dot{\bar C}\,\bar C\,C + {\bar B}^2\,\bar C\Big].
\end{eqnarray}
As a consequence, the action integrals: $S_1 = \int_{-\infty}^{\infty} d\,\tau\,L_B$ and $S_2 = \int_{-\infty}^{\infty} d\,\tau\,
L_{\bar B}$ respect the continuous and nilpotent symmetries $s_b$ and $s_{ab}$ because of the Gauss's divergence theorem 
(where all the physical variables of our theory
vanish-off as $\tau \longrightarrow  \pm\,\infty$). We can {\it also} apply $s_b$ on $L_{\bar B}$ and $s_{ab}$ on $L_B$. The ensuing results are
as follows
\begin{eqnarray}
s_b\,L_{\bar B} &=& \frac{d}{d\,\tau}\,\Big[C\,L_f - e^2\,(i\,\dot{\bar C}\,\dot C\, C + \bar B \,\dot C) - e\,\dot e\,(i\,\bar C
\,\dot C  \, C + \bar B\,C) \nonumber\\ 
&+& i\,(2\,\bar B - B)\,\bar C\,\dot C\,C + {\bar B}^2\,C\Big] \nonumber\\
&+& \big[B+ \bar B + i\,(\bar C\,\dot C - \dot{\bar C}\,C)\big]\,\big[i\,\bar C\,\ddot C\,C + e\,\dot e\,\dot C - 2\,\bar B\,\dot C + 2\,i\,
\dot{\bar C}\,\dot C \, C \big] \nonumber\\
&+& \frac{d}{d\,\tau}\big[B+ \bar B + i\,(\bar C\,\dot C - \dot{\bar C}\,C)\big]\,(e^2\,\dot C - \bar B\,C),
\end{eqnarray}
\begin{eqnarray}
s_{ab}\,L_{B} &=& \frac{d}{d\,\tau}\,\Big[\bar C\,L_f + e^2\,(i\,\dot{\bar C}\,\bar C\,\dot C + B \,\dot{\bar C}) 
+ e\,\dot e\,(i\,\dot{\bar C}\,\bar C\,C + B\,\bar C) \nonumber\\ 
&+& i\,(2\,B - \bar B)\,\dot{\bar C}\,\bar C\,C + {B}^2\,\bar C\Big] \nonumber\\
&+& \big[B+ \bar B + i\,(\bar C\,\dot C - \dot{\bar C}\,C)\big]\,(i\,\ddot{\bar C}\,\bar C\,C 
- e\,\dot e\,\dot{\bar C} - 2\,B\,\dot{\bar C} + 2\,i\,\dot{\bar C}\,\bar C\,\dot C) \nonumber\\
&-& \frac{d}{d\,\tau}\big[B+ \bar B + i\,(\bar C\,\dot C - \dot{\bar C}\,C)\big]\,(e^2\,\dot{\bar C} + B\,\bar C),
\end{eqnarray}
which demonstrate that the coupled Lagrangians $L_B$ and $L_{\bar B}$ of our theory are {\it equivalent} in the sense that {\it both} of them
respect {\it both} (i.e. BRST and anti-BRST) transformations due to the validity of the {\it physical} CF-type restriction: $B + \bar B 
+ i\,(\bar C\,\dot C - \dot{\bar C}\,C) = 0$.

It is very interesting to point out that the contributions of the term $``\chi\,\psi_5"$ in Eq. (23)
turn out to be total derivatives because we observe that:
\begin{eqnarray}
s_b\,s_{ab}\,(\chi\,\psi_5) &=& \frac{d}{d\,\tau}\,\Big[(i\,B\,\chi- {\bar C}\,\dot C\,{\chi}
- \bar C\,C\,\dot \chi)\,\psi_5 - \bar C\, C\,\chi\,\dot \psi_5\Big],\nonumber\\
-\,s_{ab}\,s_{b}\,(\chi\,\psi_5) &=& -\,\frac{d}{d\,\tau}\,\Big[(i\,\bar B\,\chi +  \dot{\bar C}\, C\,{\chi} 
+  \bar C\,C\,\dot \chi)\,\psi_5 +  \bar C\, C\,\chi\,\dot \psi_5\Big].
\end{eqnarray}
As a consequence, the gauge-fixing and Faddeev-Popov ghost terms of the Lagrangians $L_B$ and $L_{\bar B}$ 
of Eq. (24) originate from the same terms as the {\it ones} derived in our earlier work [6] on the massless spinning relativistic
particle. Thus, we note that the variable (i.e. $\psi_5)$, corresponding to the {\it mass} term for a 
{\it massive} spinning relativistic particle, does {\it not} contribute anything {\it new} to the gauge-fixing and Faddeev-Popov ghost terms.
In other words, the dynamics of our theory (at the BRST {\it quantized} level) is unaffected by the presence of the 
$``\chi\,\psi_5"$ term. This is a {\it novel} observation in our theory which is radically different from our earlier work [25] where the $``\chi\,\psi_5"$
term contributes to the dynamics. 
The observations in Eq. (30) also imply that the absolute anticommutativity property $\{s_b, s_{ab}\}\, (\chi\,\psi_5) = i\,\frac {d}{d\tau}
[B + \bar B + i\,(\bar C\,\dot C - \dot {\bar C}\,C)]$ of the (anti-)BRST symmetries $(s_{(a)b})$ is satisfied (i.e. $\{s_b, s_{ab}\} = 0 $) only when the CF-type restriction is imposed 
from {\it outside}.

According to Noether's theorem, the continuous symmetry invariance of the action integrals, corresponding to the
transformations (26) and (27), leads to the derivation of the conserved currents (i.e. conserved charges for our 1D system)  as: 
\begin{eqnarray}
J_B =  i\,B\,\bar C\,C\,\dot C + B^2\,C + B\,e^2\,\dot C + B\,e\,\dot e\,C + \frac{1}{2}\,e\,C\,(p^2 - m^2) + i\,\chi\,C\,(p_\mu\,\psi^\mu - m\,
\psi_5),
\end{eqnarray}
\begin{eqnarray}
J_{\bar B} =  i\,\bar B\,\bar C\,\dot{\bar C}\,C + {\bar B}^2\,\bar C - \bar B\,e^2\,\dot{\bar C} - \bar B\,e\,\dot e\,\bar C 
+ \frac{1}{2}\,e\,\bar C\,(p^2 - m^2) + i\,\chi\,\bar C\,(p_\mu\,\psi^\mu - m\,
\psi_5).
\end{eqnarray}
The conservation law $({dJ_r}/{d\tau}) = 0$ (with $r = B, \bar B$) can be proven by using the EL-EOMs that derived from the Lagrangians $L_B$ and $L _{\bar B}$.
For instance, we point out that the following EL-EOMs w.r.t. the variables 
($x_\mu, p_\mu, \psi_\mu, \psi_5, \chi, e, B, \bar B, C, \bar C$), namely;
\begin{eqnarray}
&& {\dot p}_\mu = 0, \qquad {\dot x}_\mu = e\,p_\mu - i\,\chi\,\psi_\mu, \qquad \psi_\mu = \chi\,p_\mu, \qquad {\dot\psi}_5 = m\,\chi, 
 \nonumber\\ 
&& p_\mu\,\psi^\mu = m\,\psi_5,  \qquad
 B - i\,(2\,\dot{\bar C}\,C + i\,e\,\dot e + \bar C\,\dot C) = 0, \nonumber\\
&& e\,\dot B + i\,e\,\dot{\bar C}\,\dot C - i\,e\,\ddot{\bar C}\,C + \frac{1}{2}\,(p^2 - m^2) = 0, \nonumber \\
&& i\,\dot B\,\bar C -i\,B\,\dot{\bar C} + i\,e\,\dot e\,\dot{\bar C} + i\,e^2\,\ddot{\bar C} + \bar C\,\ddot{\bar C}\,C + 2\,\bar C\,\dot{\bar C}
\,\dot C = 0, \nonumber\\
&& -\,i\,B\,\dot C - 2\,i\,\dot B\,C - 3\,i\,e\,\dot e\,\dot C - i\,e^2\,\ddot C - i\,{\dot e}^2\,C - i\,e\,\ddot e\,C + \bar C\,C\,\ddot C
+ 2\,\dot{\bar C}\,C\,\dot C = 0,
\end{eqnarray}
are obtained from $L_B$. The equations of motion that are different from (33) and emerge out from $L _{\bar B}$ (as the EL-EOMs) are as follows: 
\begin{eqnarray}
&& \bar B + i\,(2\,{\bar C}\,\dot C + i\,e\,\dot e + \dot{\bar C}\,C) = 0, \quad -\,e\,\dot{\bar B} + i\,e\,\dot{\bar C}\,\dot C - i\,e\,{\bar C}\,\ddot C + \frac{1}{2}\,(p^2 - m^2) = 0, \nonumber \\
&& i\,\dot{\bar B}\, C -i\,\bar B\,\dot{C} - i\,e\,\dot e\,\dot{C} - i\,e^2\,\ddot{C} + \bar C\,C\,\ddot{C} + 2\,\dot{\bar C}\,C\,\dot{C} = 0,
\nonumber\\
&& -\,i\,\bar B\,\dot{\bar C} - 2\,i\,\dot{\bar B}\,\bar C + 3\,i\,e\,\dot e\,\dot{\bar C} + i\,e^2\,\ddot{\bar C} + i\,{\dot e}^2\,\bar C 
+ i\,e\,\ddot e\,\bar C + \bar C\,\ddot{\bar C}\,C
+ 2\,{\bar C}\,\dot{\bar C}\,\dot C = 0.
\end{eqnarray}
These conserved currents $(J_B, J_{\bar B})$ lead to the definition of the conserved
 charges $Q_B$ and $Q_{\bar B}$ which are {\it same} (i.e. $J_B = Q_B, \; J_{\bar B} = Q_{\bar B}$)
as the conserved currents quoted in Eqs. (31) and (32). This is due to the fact that we are dealing with a 1D system.

We close this section with the following key comments. First, we observe that the off-shell nilpotent (anti-)BRST symmetry transformations
(20) and (21) are absolutely anticommuting in nature provided we invoke the (anti-)BRST invariant CF-type restriction: $B + \bar B +i\,(\bar C
\,\dot C- \dot{\bar C}\,C) = 0$ from {\it outside}. Second, this restriction can be obtained from the coupled (but equivalent) Lagrangians
$L_B$ and $L_{\bar B}$ if we use the EL-EOMs [cf. Eq. (25)] w.r.t. the Nakanishi-Lautrup type auxiliary variables $B$ and $\bar B$. 
Third, we observe  that the term $(\chi\,\psi_5)$ in the square bracket of Eq. (23) does {\it not} 
contribute anything to the dynamics as well as to the gauge-fixing and Faddeev-Popov
ghost terms. Fourth, the coupled Lagrangians $L_B$ and $L_{\bar B}$ are {\it equivalent} in the sense that {\it both} of them respect
{\it both} off-shell nilpotent (anti-)BRST symmetries on a submanifold of the {\it quantum} variables where the CF-type constraint:
 $B + \bar B +i\,(\bar C\,\dot C- \dot{\bar C}\,C) = 0$ is satisfied. This key observation is an {\it alternative} proof of the existence
 of CF-type restriction on our theory. Finally, we observe that the absolute anticommutativity (i.e. $\{Q_B, Q_{\bar B}\} = 0$) of the
 conserved (i.e. ${\dot Q}_{(\bar B)B} = 0$) and off-shell nilpotent (i.e. $Q_{(\bar B)B}^2 = 0$) (anti-)BRST charges  $(Q_{(\bar B) B})$ 
is satisfied only due to the validity of the {\it existence} of the CF-type restriction (cf. Sec. 6 below).

\section{Off-Shell Nilpotent Symmetries of the Target Space Variables and CF-Type Restriction: MBTSA}

In this section, we derive the off-shell nilpotent (anti-)BRST symmetry transformations for the target space variables
$ (x_\mu,\,p_\mu,\,\psi_\mu,\,\psi_5)$ which are {\it scalars} w.r.t. the 1D space of the trajectory for the
massive spinning relativistic particle that is embedded in the D-dimensional target space. For this purpose,
we exploit the theoretical power and potential of MBTSA (see e.g. [22, 26]). In this context, first of
all, we {\it promote} the 1D diffeomorphism transformation $\tau\,\longrightarrow \tau^{'} = f\,(\tau) \equiv \tau 
- \epsilon\,(\tau)$ [where $ \epsilon\,(\tau)$ is the infinitesimal transformation parameter  corresponding to the
1D diffeomorphism (i.e. reparameterization) symmetry transformation] to its counterpart on the (1, 2)-dimensional supermanifold as
\begin{eqnarray}
f\,(\tau)\,\longrightarrow f\,(\tau,\,\theta,\,\bar \theta)= \tau - \theta\,\bar C - \bar \theta \, C + \theta \,\bar \theta \,h\,(\tau),
\end{eqnarray}
where the general (1, 2)-dimensional supermanifold is parameterized by a bosonic (i.e. evolution) coordinate $ \tau$ and a pair of
Grassmannian variables $(\theta, \bar\theta)$ that satisfy: $\theta^2 = \bar\theta^2 = 0, \, \theta\,\bar\theta 
+ \bar\theta\,\theta = 0$. In the above diffeomorphism transformation, the function  $f (\tau)$ 
is any arbitrary function of the evolution parameter $\tau$  such that it is {\it finite} at $\tau = 0$ 
and vanishes-off  at $\tau  = \pm \infty$.
In other words, $f(\tau)$ is a {\it physically} well-defined function of $\tau$.
It is worth pointing out that the coefficients of the Grassmannian variables $(\theta, \bar\theta)$,
 in Eq. (35), are nothing but the fermionic
(i.e. $C^2 = {\bar C}^2 = 0, \, C\,\bar C + \bar C\,C = 0$) (anti-)ghost variables $(\bar C)\,C$ of the (anti-)BRST symmetry transformations
(20) and (21) corresponding to the infinitesimal reparameterization symmetry transformations (2). In other words, the infinitesimal 
reparameterization bosonic transformation parameter $\epsilon(\tau)$ has been replaced by the fermionic (anti-)ghost variables $(\bar C)C$ of 
the (anti-)BRST symmetry transformations. This has been done purposely in view of the fact that in earlier works
 (see e.g. [10-12]), it has
been established that the translational generators $(\partial_\theta, \partial_{\bar\theta})$, along the Grassmannian directions $(\theta,
\bar\theta)$, are intimately connected with the nilpotent (anti-)BRST transformations $s_{(a)b}$ in the {\it ordinary} space. In other words,
we have already taken into account $s_{ab}\,\tau = - \,\bar C, s_b\,\tau = -\,C$ which are the generalization of the {\it classical} infinitesimal
1D diffeomorphism symmetry transformation: $\delta_r\,\tau = -\,\epsilon(\tau)$ to its {\it quantum} counterparts
 $(s_{(a)\,b})$ within the framework of  BRST formalism.  It is worthwhile to point out that the secondary variable $h(\tau)$ of
the expansion (37) has to be determined from {\it other} consistency conditions which we elaborate in our forthcoming paragraphs.

According to the basic tenets of MBTSA, we have to {\it generalize} {\it all} the ordinary variables of the Lagrangians (24) onto the (1, 2)-dimensional
supermanifold as {\it their} counterparts supervariables where the {\it generalization} of the 1D diffeomorphism transformation
[cf. Eq. (35)] has to be
incorporated (in a suitable fashion) into the expressions for the supervariables. For instance, we shall have the following generalization
as far as
the {\it generic} target space variable $s_\mu(\tau)$ [cf. Eq. (22)] is concerned, namely;
\begin{eqnarray}
s_\mu(\tau) \quad \longrightarrow \quad {\tilde S}_\mu [f(\tau, \theta, \bar\theta), \theta, \bar\theta] \equiv {\tilde S}_\mu
[\tau 
- \theta\,\bar C - \bar\theta\,C + \theta\,\bar\theta\,h, \theta, \bar\theta], 
\end{eqnarray}
where the pair of variables $(\theta, \bar\theta)$, as pointed out earlier, are the Grassmannian variables (i.e. $\theta^2 = {\bar\theta}^2 = 0,
\; \theta\,\bar\theta + \bar\theta\,\theta = 0$) of the superspace coordinates $Z^M = (\tau, \theta, \bar\theta)$ that characterize 
the (1, 2)-dimensional supermanifold on which our 1D {\it ordinary} theory of the reparameterization invariant massive spinning particle is considered. 
Now, following the
techniques of MBTSA, we have the following super expansion of (36) along {\it all} the possible directions of the Grassmannian variables of 
the $(1, 2)$-dimensional supermanifold, namely;
\begin{eqnarray}
{\tilde S}_\mu[f(\tau, \theta, \bar\theta), \theta, \bar\theta] = S_\mu[f(\tau, \theta, \bar\theta)] + \theta\,{\bar R}_\mu
[f(\tau, \theta, \bar\theta)] + \bar\theta\,R_\mu [f(\tau, \theta, \bar\theta)] + \theta\,\bar\theta\,P_\mu[f(\tau, \theta, \bar\theta)], 
\end{eqnarray}
where, on the r.h.s., we have the {\it secondary} supervariables which are {\it also} function of $f(\tau, \theta, \bar\theta)$. As a
consequence, we can have the following Taylor expansions (for those secondary variables that are present on the r.h.s.), namely;
\begin{eqnarray}
&&S_\mu\,(\tau - \theta \, \bar C - \bar \theta \, C + \theta \, \bar \theta \, h) = s_\mu \, (\tau) 
- \theta\, \bar C \, {\dot s_\mu \,(\tau) - \bar \theta \, C \dot s_\mu \, (\tau)
+ \theta \bar \theta\,(h\,\dot s_\mu - \bar C\, C\, \Ddot s_\mu)},\nonumber\\
&&\theta \, \bar R_\mu \,(\tau - \theta \, \bar C - \bar \theta \, C + \theta \, \bar \theta \, h ) 
\, \equiv \, \theta \, \bar R_\mu(\tau) - \theta \bar \theta \, \dot{ \bar {R_\mu }}\,(\tau), \nonumber\\
&&\bar \theta \,  R_\mu \,(\tau - \theta \, \bar C - \bar \theta \, C + \theta \, \bar \theta \, h ) 
\,\equiv \, \bar \theta \,  R_\mu (\tau) + \theta \bar \theta \, \dot{R_\mu}\,(\tau),\nonumber\\
&&\theta \,\bar \theta\, P_\mu \,(\tau - \theta \, \bar C - \bar \theta \, C + \theta \, \bar \theta \, h )
\, \equiv \, \theta \,\bar \theta \, P_\mu(\tau).
\end{eqnarray}
At this juncture, we would like to lay stress on the fact that, in the super expansion (37),
{\it all} the supervariables on the r.h.s. have to be {\it ordinary} variables as {\it all} of them
are Lorentz {\it scalars} w.r.t. the 1D trajectory of the particle (that is embedded in the D-dimensional 
flat Minkowskian target space). It is worthwhile to point out that a pure Lorentz (bosonic or fermionic) scalar is the one which does {\it not} transform {\it at all}
under any kind of {\it spacetime} and/or {\it internal} transformations. As a result, the expansion (37) can be written as:
\begin{eqnarray}
\tilde S_\mu\,[f\,(\tau,\,\theta,\,\bar\theta),\,\theta,\,\bar\theta] & = & s_\mu\,(\tau) 
+ \theta \, \bar R_\mu\,(\tau) + \bar\theta \,  R_\mu\,(\tau) + \theta \,\bar\theta \, P_\mu\,(\tau)\nonumber\\
& \equiv & s_\mu\,(\tau) + \theta \,(s_{ab}\,s_\mu\,(\tau)) + \bar \theta \,(s_{b}\,s_\mu\,(\tau))
+ \theta \bar \theta \,(s_b\,s_{ab}\,s_\mu\,(\tau)),
\end{eqnarray}
where $ s_{(a)b}$ are the (anti-)BRST symmetry transformations (20) and (21). This is due to fact that the (anti-)BRST
symmetry transformations $s_{(a)b}$ have been shown to be deeply connected with the translational generators 
$(\partial_\theta, \partial_{\bar\theta})$ along the $(\theta, \bar\theta)$-directions of the (1, 2)-dimensional 
supermanifold (see e.g. [10-12] for details).

It is evident  that we have to compute the values of $R_\mu, \bar R_\mu$ and $P_\mu$ [in terms of the auxiliary and 
basic variables of the Lagrangians (24)] so that we can obtain the off-shell nilpotent (anti-)BRST symmetry transformations
for the {\it generic} variable $s_\mu (\tau)$. At this stage, the so called ``horizontality condition" (HC)
comes to our help where we demand that: $\tilde S_\mu (f(\tau, \theta, \bar\theta), \theta, \bar\theta) = s_\mu (\tau).$
This relationship can be explicitly expressed as 
\begin{eqnarray}
&&s_\mu (\tau) + \theta\,(\bar R_\mu - \bar C\,\dot s_\mu) + \bar\theta\,(R_\mu - C\,\dot s_\mu) \nonumber\\
&&+ \theta\,\bar\theta\,
[h\,\dot s_\mu  - \bar C\,C\,\ddot x_\mu - C\,\dot{\bar {R_\mu}} + \bar C\,\dot R_\mu + P_\mu] 
\equiv s_\mu (\tau),
\end{eqnarray}
where we have collected all the terms from Eq. (38) to express (36). Physically, the above requirement corresponds 
to the fact that a Lorentz-scalar does {\it not} transform under any kind of physically well-defined spacetime transformations .
Needless to say, the relationship (40) implies that we have the following explicit relationships:
\begin{eqnarray}
R_\mu = C\,\dot s_\mu , \quad \bar R_\mu = \bar C\,{\dot s}_\mu, \quad P_\mu = C\, \dot {\bar {R_\mu}}
- \bar C\,\dot R_\mu + \bar C\, C \, \dot s_\mu  - h\,\ddot s_\mu.
\end{eqnarray}
It is straightforward to note that we have already obtained $s_b\,s_\mu = C\,{\dot s}_\mu$ and $s_{ab}\,s_\mu = \bar C\,{\dot s}_\mu$ as is
evident from Eq. (39). The requirement of the absolute anticommutativity [that is {\it one} of the sacrosanct properties of the (anti-)BRST
symmetry transformations] implies that we have the following equalities, namely; 
\begin{eqnarray}
s_b\,s_{ab}\,s_\mu = -\,s_{ab}\,s_b\,s_\mu \quad \implies \quad \{s_b, s_{ab}\}\,s_\mu = 0.
\end{eqnarray}
On the other hand, the requirement of the off-shell nilpotency [that is {\it another} sacrosanct property of the (anti-)BRST symmetry
transformations] leads to the following:
\begin{eqnarray}
s_b\,C = C\,\dot C, \qquad \qquad s_{ab}\,\bar C = \bar C\,\dot{\bar C}.
\end{eqnarray}
On top of the already obtained off-shell nilpotent (anti-)BRST symmetry transformations: 
$s_b\,s_\mu = C\,{\dot s}_\mu, \, s_{ab}\,s_\mu = \bar C\,{\dot s}_\mu, \,
s_b\,C = C\,\dot C, \, s_{ab}\,\bar C = \bar C\,\dot{\bar C},$ we take into account the standard (anti-)BRST symmetry transformations $s_b\,\bar 
C = i\,B$ and $s_{ab}\, C = i\,\bar B$ in terms of the Nakanishi-Lautrup auxiliary variables. These standard inputs lead to the determination of
the l.h.s. and r.h.s. of the {\it first} equality in Eq. (42) as [22, 26]:
\begin{eqnarray}
s_b\,s_{ab}\,s_\mu = (i\,B - \bar C\,\dot C)\,{\dot s}_\mu - \bar C\,C\,{\ddot s}_\mu \equiv P_\mu(\tau), \nonumber \\
-\,s_{ab}\,s_{b}\,s_\mu = (- i\,\bar B - \dot{\bar C}\,C)\,{\dot s}_\mu - \bar C\,C\,{\ddot s}_\mu \equiv P_\mu(\tau),
\end{eqnarray}
where $P_\mu\,(\tau)$ is present in the expansion (39). A close look at Eq. (44) implies that we have: $B + \bar B + i\,(\bar C\,\dot C - \dot{\bar C}\,C) = 0$ which is nothing but the CF-type
restriction. In addition, the observation of Eq. (41) implies that there is an explicit expression for $P_\mu$ in terms of $h(\tau)$ 
[that is present in the expansion of $f(\tau, \theta, \bar\theta)$ in Eq. (35)]. 
Plugging in the values of $ R_\mu = C \,\dot s_\mu$ and  $ \bar R_\mu = \bar C \,\dot s_\mu$ into Eq. (41) leads to [22, 26] 
\begin{eqnarray}
&&P_\mu \,(\tau) = -\,[ (\dot {\bar C}\,C + \bar C \, \dot C + h)\,\dot s_\mu + \bar C \,C\, \ddot s_\mu].
\end{eqnarray}
Comparison of (44) and (45) yields (see, e.g. [22, 26] for details):
\begin{eqnarray}
&&h = -\,i\,B - \dot{\bar C}\,C \,\equiv \, +\,i\,\bar B - \bar C\, \dot C \, \implies \,  B 
+ \bar B + i\,(\bar C \, \dot C + \dot{\bar C}\, C) = 0.
\end{eqnarray}
Thus, we note that the comparison of the values of $ h\,(\tau)$ [that is determined from the comparison
between Eq. (44) and Eq. (45)] leads to the deduction of the (anti-)BRST invariant (i.e. $s_{(a)b}\,
[B + \bar B + i\,(\bar C\,\dot C - \dot{\bar C}\,C)] = 0$) CF-type restriction: $ B + \bar B + i\,(\bar C \, \dot C
- \dot{\bar C}\, C) = 0$ which plays an important role in the proof: $\{s_b, s_{ab}\} = 0$.

We wrap-up this section with the following useful and important remarks. First, we have taken 
into account the standard choice in the BRST formalism which is: $ s_b\,\bar C = i\,B, \, s_{ab}\,C = i\, \bar B$. 
In other words, we have made the following (anti-)chiral super expansions for the (anti-)chiral supervariables 
(in view of $\partial_{\bar \theta} \leftrightarrow s_{ab},\,\partial_{\theta} \leftrightarrow s_{b},$), namely;
\begin{eqnarray}
&& C\,(\tau)\,\longrightarrow\, F^{(c)}\,(\tau,\,\theta) = \, C\,(\tau) + \theta \, [i\, \bar B(\tau)] \equiv C\,(\tau) 
+ \theta\, [s_{ab}\,C(\tau)],\nonumber\\
&& \bar C\,(\tau)\,\longrightarrow\, \bar F^{(ac)}\,(\tau,\,\bar \theta) = \, \bar C\,(\tau) + \bar \theta \, 
[i\, B(\tau)] \equiv \bar C\,(\tau) + \bar \theta\, [s_{b}\,\bar C(\tau)],
\end{eqnarray}
where $F^{(c)}\,(\tau,\,\theta)$ and $ \bar F^{(ac)}\,(\tau,\,\bar \theta) $ are the {\it chiral} 
and anti-chiral supervariables that have been defined on the (1, 1)-dimensional suitably chosen {\it chiral} and 
{\it anti-chiral} super sub-manifolds of the {\it general} (1, 2)-dimensional supermanifold.
Second, we have seen that [cf. Eq. (22)] the absolute anticommutativity property (i.e. $ \{s_b, s_{ab}\} = 0$)
of the (anti-)BRST transformations $(s_{(a)b}) $ is satisfied if and only if the CF-type restriction
(46) is satisfied. Third, we point out that the requirement of the following 
\begin{eqnarray}
&&\{s_b, s_{ab}\}\,C = 0\; \implies\; s_{b}\,{\bar B} =  \dot{\bar B}\,{C} - \bar B\,\dot{C}, \nonumber\\
&&\{s_b, s_{ab}\}\,{\bar C} = 0\; \implies \; s_{ab}\,{ B} =  \dot{B}\,{\bar C} -  B\,\dot{\bar C}.
\end{eqnarray}
leads to the derivation of (anti-)BRST symmetry transformations on the Nakanishi-Lautrup auxiliary
variables $(B)\bar B$. Fourth, within the framework of MBTSA, the CF-type restriction (46) is derived
from the expression for $h(\tau)$ due to the consistency condition (i.e. $s_b s_{ab}\,
s_\mu = - \,s_{ab} s_b\, s_\mu \equiv P_\mu$). Fifth, the (anti-)BRST symmetry transformations ($ s_{b} s_\mu = {C}\, 
{\dot s}_\mu,\,\,s_{ab} s_\mu = {\bar C}\, {\dot s}_\mu$) on the generic variable $s_\mu \equiv 
x_\mu,\,p_\mu,\,\psi_\mu,\,\psi_5$ imply that we have already obtained the following (anti-)BRST symmetry transformations
\begin{eqnarray}
&&s_{ab}\; x_\mu = {\bar C}\; {\dot x}_\mu, \qquad s_{ab}\; p_\mu = {\bar C}\; {\dot p}_\mu,\qquad
s_{ab} \;\psi_\mu = {\bar C}\; {\dot\psi}_\mu, \qquad s_{ab} \,\psi _5 = {\bar C}\; {\dot\psi}_5, \nonumber\\
&&s_{b}\; x_\mu = {C}\; {\dot x}_\mu, \qquad\; s_{b}\; p_\mu = {C}\; {\dot p}_\mu, \qquad\; s_{b} \;\psi_\mu =
{C}\; {\dot\psi}_\mu,\qquad s_{b} \,\psi _5 = { C}\; {\dot\psi}_5, 
\end{eqnarray}
for the target space variables $(x_\mu, p_\mu, \psi_\mu, \psi_5)$ of our theory that are present in the first-order 
Lagrangian $L_f$ [cf. Eq. (1)] for the massive spinning  (i.e. SUSY) relativistic particle. Finally, the explicit form of Eq. (39) 
can be written, after the application of HC, as follows
\begin{eqnarray}
X_\mu^{(h)} (f(\tau, \theta, \bar\theta), \theta, \bar\theta) & = & x_\mu(\tau) + \theta\,(\bar C\,{\dot x}_\mu) + \bar\theta\,(C\,{\dot x}_\mu)
+ \theta\,\bar\theta\,[(i\,B - \bar C\,\dot C)\,{\dot x}_\mu - \bar C\,C\,{\ddot x}_\mu],\nonumber\\
& \equiv & x_\mu(\tau) + \theta\,(s_{ab}\,x_\mu) + \bar\theta\,(s_b\,x_\mu)
+ \theta\,\bar\theta\,(s_b\,s_{ab}\,x_\mu),\nonumber\\
P_\mu^{(h)} (f(\tau, \theta, \bar\theta), \theta, \bar\theta) & = & p_\mu(\tau) + \theta\,(\bar C\,{\dot p}_\mu) + \bar\theta\,(C\,{\dot p}_\mu)
+ \theta\,\bar\theta\,[(i\,B - \bar C\,\dot C)\,{\dot p}_\mu - \bar C\,C\,{\ddot p}_\mu], \nonumber\\
& \equiv & p_\mu(\tau) + \theta\,(s_{ab}\,p_\mu) + \bar\theta\,(s_b\,p_\mu)
+ \theta\,\bar\theta\,(s_b\,s_{ab}\,p_\mu),\nonumber\\
\Psi_\mu^{(h)} (f(\tau, \theta, \bar\theta), \theta, \bar\theta) & = & \psi_\mu(\tau) + \theta\,(\bar C\,{\dot \psi}_\mu) + \bar\theta\,(C\,{
\dot \psi}_\mu) + \theta\,\bar\theta\,[(i\,B - \bar C\,\dot C)\,{\dot \psi}_\mu - \bar C\,C\,{\ddot \psi}_\mu], \nonumber\\
& \equiv & \psi_\mu(\tau) + \theta\,(s_{ab}\,\psi_\mu) + \bar\theta\,(s_b\,\psi_\mu)
+ \theta\,\bar\theta\,(s_b\,s_{ab}\,\psi_\mu),\nonumber\\
\Psi_5^{(h)} (f(\tau, \theta, \bar\theta), \theta, \bar\theta) & = & \psi_5(\tau) + \theta\,(\bar C\,{\dot \psi}_5) + \bar\theta\,(C\,{
\dot \psi}_5) + \theta\,\bar\theta\,[(i\,B - \bar C\,\dot C)\,{\dot \psi}_5 - \bar C\,C\,{\ddot \psi}_5], \nonumber\\
& \equiv & \psi_5(\tau) + \theta\,(s_{ab}\,\psi_5) + \bar\theta\,(s_b\,\psi_5)
+ \theta\,\bar\theta\,(s_b\,s_{ab}\,\psi_5)
\end{eqnarray}
where the superscript $(h)$ denotes the full expansion of the supervariables after the application of HC. A straightforward comparison of
(39) with (50) shows that we have already derived the (anti-)BRST symmetry transformations (49) as the coefficients of $(\theta)\bar\theta$
in the super expansions (50) along with $ s_b\,s_{ab}\,x_\mu,  s_b\,s_{ab}\,\psi_\mu,  s_b\,s_{ab}\,p_\mu,  s_b\,s_{ab}\,\psi_5$ which are
the coefficients of $\theta \bar\theta$.
 We also note, from Eq. (50), that we have a mapping: $s_b\, \leftrightarrow \, \partial_{\bar \theta}|_{\theta = 0},\, 
s_{ab}\, \leftrightarrow \, \partial_{\theta}|_{\bar\theta= 0}$. This observation is consistent with results 
obtained  in  the Refs. [10-12].

\section{Off-Shell Nilpotent (Anti-)BRST Symmetries for Other Variables of Our Theory: ACSA}

In this section, we exploit the basic principle behind ACSA to BRST formalism to obtain,
 first of all, the off-shell nilpotent BRST symmetry transformations
(21) by generalizing the basic and auxiliary variables of the Lagrangian $L_B$ [cf. Eq. (24)]
 on the {\it anti-chiral} $(1, 1)$-dimensional super sub-manifold 
[of the most {\it general} (1, 2)-dimensional supermanifold] as 
\begin{eqnarray}
B(\tau) \quad &\longrightarrow& \quad \tilde B (\tau, \bar\theta) = B(\tau) + \bar\theta\,f_1(\tau), \nonumber\\
e(\tau) \quad &\longrightarrow& \quad  E(\tau, \bar\theta) = e(\tau) + \bar\theta\,f_2(\tau), \nonumber\\
\chi(\tau) \quad &\longrightarrow& \quad  K(\tau, \bar\theta) = \chi(\tau) + \bar\theta\,b_1(\tau), \nonumber\\
C(\tau) \quad &\longrightarrow& \quad  F(\tau, \bar\theta) = C(\tau) + \bar\theta\,b_2(\tau), \nonumber\\
\bar C(\tau) \quad &\longrightarrow& \quad  \bar F(\tau, \bar\theta) = \bar C(\tau) + \bar\theta\,b_3(\tau), \nonumber\\
\bar B(\tau) \quad &\longrightarrow& \quad  \tilde{\bar B}(\tau, \bar\theta) = \bar B(\tau) + \bar\theta\,f_3(\tau), 
\end{eqnarray}
where $(f_1, f_2, f_3)$ are the fermionic and $(b_1, b_2, b_3)$ are bosonic {\it secondary} variables.
 These {\it variables} have to be expressed in terms of the auxiliary and basic variables 
 that are present in $L_B$. For this purpose, we exploit the BRST-invariant restrictions. 
It will be noted that the anti-chiral 
(1, 1)-dimensional super sub-manifold is parameterized by the superspace 
coordinates $Z^{M} = (\tau,\, \bar \theta)$ where $\tau$ is the {\it bosonic}
 evolution parameter and Grassmannian variable $\bar \theta$ is fermionic 
($ {\bar \theta}^2 = 0$) in nature. In addition to (51), we have the {\it 
anti-chiral} limit (i.e. $ \theta = 0$) of the expansions (50) as follows
\begin{eqnarray}
&&X_\mu^{(ha)}(\tau, \bar\theta) = x_\mu(\tau) + \bar\theta\,(C\,{\dot x}_\mu), \quad\qquad \Psi_\mu^{(ha)}(\tau, \bar\theta) = \psi_\mu(\tau) 
+ \bar\theta\,(C\,{\dot\psi}_\mu), \nonumber\\
&&P_\mu^{(ha)}(\tau, \bar\theta) = p_\mu(\tau) + \bar\theta\,(C\,{\dot p}_\mu), ~\quad\qquad \Psi_5^{(ha)}(\tau, \bar\theta) = \psi_5(\tau) + 
 \bar\theta\,(C\,{\dot\psi}_5),
\end{eqnarray} 
where the superscript $(ha)$ denotes the anti-chiral limit of the super expansions (of the supervariables [cf. Eq. (50)]) that have been obtained after 
the application of HC. It is straightforward to note that the BRST invariance ($s_b\,B = 0$) of the variable $B$ implies that we have the following (with $f_1(\tau) = 0$), namely;
\begin{eqnarray}
{\tilde B}(\tau, \bar\theta) \quad \longrightarrow  \quad {\tilde B}^{(b)}(\tau, \bar\theta)
 = B(\tau) + \bar\theta\,(0) = B(\tau) + \bar\theta\,(s_b\, B),
\end{eqnarray}  
where the superscript $(b)$ stands for the anti-chiral supervariable that has been obtained after the BRST invariant $(s_b\,B = 0)$ restriction. In other
words, we have already obtained the BRST symmetry transformation $s_b\,B = 0$ as the coefficient of $\bar\theta$ in (53) due to our knowledge of:
$s_b \leftrightarrow \partial_{\bar\theta}$ [i.e. $\partial_{\bar\theta}\,{\tilde B}^{(b)}(\tau, \bar\theta) = s_b\,B = 0$].

The off-shell nilpotency of the BRST symmetry transformations (21) ensures that we have the following BRST invariant quantities:   
\begin{eqnarray}
&&s_b\,(C\,{\dot x}_\mu) = 0, \quad s_b\,(C\,{\dot p}_\mu) = 0, \quad s_b\,(C\,\dot C) = 0, \quad s_b\,(C\,\dot e + \dot C\, e) = 0, \nonumber\\
&&s_b\,(\dot C\, \chi + C \, \dot \chi) = 0, \quad s_b\,(C\,{\dot\psi}_\mu) = 0, \quad s_b(C\,{\dot\psi}_5)
 = 0, \quad s_b\,(\dot{\bar B}\,C - \bar B \,\dot C) = 0. 
\end{eqnarray}
The above {\it quantum} gauge (i.e. BRST) invariant quantities must be independent of the Grassmannian variable $\bar\theta$ when they are 
generalized onto $(1, 1)$-dimensional {\it anti-chiral} super sub-manifold [of the most {\it general} $(1, 2)$-dimensional supermanifold)] on which our 1D 
{\it ordinary} theory has been generalized. In other words, we have the validity of the following equalities in terms of the supervariables and ordinary variables; namely;
\begin{eqnarray}
&&F(\tau, \bar\theta)\,{\dot X}_\mu^{(ha)}(\tau, \bar\theta) = C(\tau)\,{\dot x}_\mu(\tau), \qquad\quad F(\tau, \bar\theta)\,{\dot F}(\tau, \bar\theta) 
= C(\tau)\,{\dot C}(\tau) \nonumber\\
&&F(\tau, \bar\theta)\,{\dot\Psi}_\mu^{(ha)}(\tau, \bar\theta) = C(\tau)\,{\dot\psi}_\mu(\tau), \qquad\quad F(\tau, \bar\theta)\,{\dot\Psi}_5^{(ha)}
(\tau, \bar\theta) = C(\tau)\,{\dot\psi}_\mu(\tau), \nonumber\\ 
&&F(\tau, \bar\theta)\,{\dot P}_\mu^{(ha)}(\tau, \bar\theta) = C(\tau)\,{\dot p}_\mu(\tau), \qquad\quad {\tilde B}^{(b)}(\tau, \bar\theta) = 
B(\tau), \nonumber\\ 
&&\dot F(\tau, \bar\theta)\,E(\tau, \bar\theta) + F(\tau, \bar\theta)\,\dot E(\tau, \bar\theta) = \dot C(\tau)\,e(\tau) + C(\tau)\,\dot e(\tau), \nonumber\\
&&\dot F(\tau, \bar\theta)\,K(\tau, \bar\theta) + F(\tau, \bar\theta)\,\dot K(\tau, \bar\theta) = \dot C(\tau)\,\chi(\tau) + C(\tau)\,\dot\chi(\tau),  \nonumber\\ 
&&\dot{\tilde{\bar B}}(\tau,\bar\theta)\,F(\tau, \bar\theta) - \tilde{\bar B}(\tau, \bar\theta)\,{\dot F}(\tau, \bar\theta) = \dot{\bar B}(\tau)\,
C(\tau) - {\bar B}(\tau)\,\dot C(\tau),
\end{eqnarray}
where the supervariables with superscripts $(ha)$ and $(b)$ have already been  explained and derived in Eqs. (52) and (53). 
The substitutions of the expansions 
from (52) and (51) lead to the determination of the secondary variables of the {\it latter} equation [cf. Eq. (51)] as:
\begin{eqnarray}
&&f_2(\tau) = \dot C\,e + C\,\dot e, \qquad b_1(\tau) = \dot C\,\chi + C\,\dot\chi, \qquad b_2(\tau) = C\,\dot C \nonumber\\ 
&&b_3(\tau) = i\,B, \qquad f_3(\tau) = \dot{\bar B}\,C - \bar B \,\dot C. 
\end{eqnarray}
The above relationships demonstrate that we have already obtained the secondary variables of the super expansion (51) in terms of the auxiliary and basic
 variables of 
$L_B$ (and the Nakanishi-Lautrup auxiliary variable $\bar B(\tau)$ of the Lagrangian $L_{\bar B}$ [cf. Eq. (24)]).

The substitutions of the above expressions for the secondary variables into the super expansions (51) [besides Eqs. (52), (53)] are as follows
\begin{eqnarray*}
&&E^{(b)}(\tau, \bar\theta) = e(\tau) + \bar\theta\,(e\,\dot C + \dot e\,C) \equiv e(\tau) + \bar\theta\,(s_b\,e), \nonumber\\
\end{eqnarray*}
\begin{eqnarray}
&&K^{(b)}(\tau, \bar\theta) = \chi(\tau) + \bar\theta\,(C\,\dot\chi + \dot C\,\chi) \equiv \chi(\tau) + \bar\theta\,(s_b\,\chi),\nonumber\\ 
&&F^{(b)}(\tau, \bar\theta) = C(\tau) + \bar\theta\,(C\,\dot C ) \equiv C(\tau) + \bar\theta\,(s_b\,C), \nonumber\\ 
&&{\bar F}^{(b)}(\tau, \bar\theta) = \bar C(\tau) + \bar\theta\,(i\,B) \equiv \bar C(\tau) + \bar\theta\,(s_b\,\bar C), \nonumber\\
&&{\tilde{\bar B}}^{(b)}(\tau, \bar\theta) = \bar B(\tau) + \bar\theta\,(\dot{\bar B}\,C - \bar B\,\dot C ) \equiv \bar B(\tau) + \bar\theta\,(s_b\,\bar B), 
\end{eqnarray}
where the superscript $(b)$ on the supervariables denotes the {\it anti-chiral} supervariables that have been obtained after the imposition of the
BRST (i.e. {\it quantum} gauge) invariant restrictions in Eq. (55). It is clear from (57) that we have a mapping: $\partial_{\bar\theta}
\leftrightarrow s_b$ which demonstrates that the off-shell nilpotency $(\partial_{\bar\theta}^2 = 0)$ of the translational generator
 $\partial_{\bar\theta}$ along 
$\bar\theta$-direction of the $(1, 1)$-dimensional {\it anti-chiral} super sub-manifold {\it and} off-shell nilpotency $(s_b^2 = 0)$ of the 
BRST transformations (21) in the {\it ordinary} space are interrelated. A careful look at Eqs. (50) and (57) demonstrate that we have 
already derived the BRST
symmetry transformations (21) for {\it all} the variables of $L_B$ as the coefficients of $\bar\theta$.

Now we dwell a bit on the derivation of the anti-BRST transformations (20) within the framework of ACSA. Towards this end 
in mind, we note that the following are the {\it chiral} (i.e. $ \bar \theta = 0$) limit of the full super expansions in (50), namely;
\begin{eqnarray}
&&X_\mu^{(hc)}(\tau, \theta) = x_\mu(\tau) + \theta\,(\bar C\,{\dot x}_\mu) \equiv  x_\mu(\tau) + \theta\,(s_{ab}\,x_\mu), \nonumber\\ 
&&P_\mu^{(hc)}(\tau, \theta) = p_\mu(\tau) + \theta\,(\bar C\,{\dot p}_\mu)  \equiv  p_\mu(\tau) + \theta\,(s_{ab}\,p_\mu),\nonumber\\
&&\Psi_\mu^{(hc)}(\tau, \theta) = \psi_\mu(\tau) + \theta\,(\bar C\,{\dot\psi}_\mu)  \equiv  \psi_\mu(\tau) + \theta\,(s_{ab}\,\psi_\mu), \nonumber\\
 && \Psi_5^{(hc)}(\tau, \theta) = \psi_5(\tau) +  \theta\,(\bar C\,{\dot\psi}_5)  \equiv  \psi_5(\tau) + \theta\,(s_{ab}\,\psi_5),
\end{eqnarray}
where the superscript $(hc)$ stands for the {\it chiral}  limit of the supervariables that have been derived after the application of HC in Eq. (50).
 The above expansions in (58) would be utilized in the anti-BRST invariant restrictions on the {\it chiral}  supervariables which we are going to
 discuss as follows. It can be readily checked that we have the following interesting anti-BRST invariant quantities
\begin{eqnarray}
&&s_{ab}\,(\bar C\,{\dot x}_\mu) = 0, \quad s_{ab}\,(\bar C\,{\dot p}_\mu) = 0, \quad s_{ab}\,(\bar C\,\dot {\bar C}) = 0, \quad s_{ab}\,(\dot{\bar C}\,e
+ \bar C\,\dot e) = 0, \nonumber\\
&&s_{ab}\,(\dot{\bar C}\,\chi + \bar C\,\dot\chi) = 0, \quad s_{ab}\,(\bar C\,{\dot\psi}_\mu) = 0, \quad s_{ab}(\bar C\,{\dot\psi}_5) = 0, 
\quad s_{ab}\,(\dot{B}\,\bar C - B \,\dot {\bar C}) = 0, 
\end{eqnarray}
where the anti-BRST symetry transformations $(s_{ab})$ are the {\it ones} that have been listed in Eq. (20). Keeping in our mind the 
mapping: $s_{ab} \leftrightarrow \partial_{\theta}$ and the observation $s_{ab}\,\bar B = 0$,we have the following
\begin{eqnarray}
{\tilde {\bar B}}(\tau, \theta) \quad \longrightarrow  \quad {\tilde {\bar B}}^{(ab)}(\tau, \theta) = \bar B(\tau) + \theta\,(0) 
= \bar B(\tau) + \theta\,(s_{ab}\, \bar B),
\end{eqnarray}
where the superscript $(ab)$ denotes the expansion of the supervariable that has been 
obtained after the application of the anti-BRST invariant
 restriction: $\tilde{\bar B}(\tau, \theta) = \bar B(\tau)$
that is obtained due to the anti-BRST invariance $(s_{ab}\,\bar B = 0)$. We also note that $\partial_{\theta}\,{\tilde{\bar B}}^{(ab)}
(\tau, \theta) = s_{ab}\,\bar B = 0$. Exploiting the basic principle of ACSA to BRST formalism,
 we obtain the following equalities in terms of 
{\it chiral} and {\it ordinary} variables, namely;
 \begin{eqnarray*}
 &&\bar F(\tau, \theta)\,{\dot X}_{\mu}^{(hc)}(\tau, \theta) = \bar C(\tau)\,{\dot x}_{\mu}(\tau), \qquad\quad \bar F(\tau, \theta)\,
{\dot P}_\mu^{(hc)}(\tau, \theta) = \bar C(\tau)\,{\dot p}_\mu(\tau), \nonumber\\
&&\bar F(\tau, \theta)\,{\dot\Psi}_{\mu}^{(hc)}(\tau, \theta) = \bar C(\tau)\,{\dot\psi}_{\mu}(\tau), \qquad\quad \bar F(\tau, \theta)\,
{\dot\Psi}_5^{(hc)}(\tau, \theta) = \bar C(\tau)\,{\dot\psi}_5(\tau), 
\end{eqnarray*}
\begin{eqnarray}
&&\bar F(\tau, \theta)\,{\dot{\bar F}}(\tau, \theta) = \bar C(\tau)\,\dot{\bar C}(\tau), \qquad\quad\quad\;\; {\tilde{\bar B}}^{(ab)}(\tau, \theta) 
= \bar B(\tau), \nonumber\\
&&\dot{\bar F}(\tau, \theta)\,E(\tau, \theta) + \bar F(\tau, \theta)\,\dot E(\tau, \theta) = \dot{\bar C}(\tau)\,e(\tau) + \bar C(\tau)\,\dot e(\tau),\nonumber\\
&&\dot{\bar F}(\tau, \theta)\,K(\tau, \theta) + \bar F(\tau, \theta)\,\dot K(\tau, \theta) = \dot{\bar C}(\tau)\,\chi(\tau) + \bar C(\tau)\,\dot\chi(\tau),
 \nonumber\\
 &&\dot{\tilde B}(\tau, \theta)\,\bar F(\tau, \theta) - \tilde B(\tau, \theta)\,\dot{\bar F}(\tau, \theta) = \dot B(\tau)\,\bar C(\tau) 
- B(\tau)\,\dot{\bar C}(\tau), 
 \end{eqnarray}
where the {\it chiral} supervariables with superscripts $(hc)$  and $(ab)$ have been discussed and explained in Eqs. (58) and (60). It is worth pointing 
out that the equalities in (61) are nothing but the generalization of our observations in (59) where the anti-BRST invariant quantities
have been obtained [because of the off-shell nilpotency of the anti-BRST symmetry transformations (20)]. At this stage, it is crucial to 
point out that, besides our {\it chiral} supervariables in (58) and (60), we have the following generalizations:
\begin{eqnarray}
B(\tau) \qquad &\longrightarrow& \qquad \tilde B(\tau, \theta) = B(\tau) + \theta\,{\bar f}_1(\tau), \nonumber\\
e(\tau) \qquad &\longrightarrow& \qquad E(\tau, \theta) = e(\tau) + \theta\,{\bar f}_2(\tau), \nonumber\\
\chi(\tau) \qquad &\longrightarrow& \qquad K(\tau, \theta) = \chi(\tau) + \theta\,{\bar b}_1(\tau), \nonumber\\
C(\tau) \qquad &\longrightarrow& \qquad F(\tau, \theta) = C(\tau) + \theta\,{\bar b}_2(\tau), \nonumber\\
\bar C(\tau) \qquad &\longrightarrow& \qquad \bar F(\tau, \theta) = \bar C(\tau) + \theta\,{\bar b}_3(\tau).
\end{eqnarray} 
The above {\it chiral} supervariables are defined and their expansions have been carried out on a $(1, 1)$-dimensional super sub-manifold
that is characterized by the superspace coordinates $Z^M = (\tau, \theta)$ where the Grassmannian coordinate $\theta$ is fermionic 
$(\theta^2 = 0)$ in nature. It is straightforward to draw the conclusion that the secondary variables $({\bar b}_1, {\bar b}_2, {\bar b}_3)$
and $({\bar f}_1, {\bar f}_2, {\bar f}_3)$, on the r.h.s. of Eq. (62) are bosonic and fermionic sets, respectively.

The stage is now set to utilize the equalities (61) where we have to plug in the chiral super expansions (58) as well as the chiral generalizations
(62). This exercise leads to the following relationships between the secondary variables of (62) and the basic as well as auxiliary variables of
 Lagrangian $L_{\bar B}$ (and the Nakanishi-Lautrup auxiliary variable $B$ of the perfectly  BRST invariant Lagrangian $L_B$), namely; 
\begin{eqnarray}
&&{\bar f}_1 = \dot B\,{\bar C} - B\,\dot{\bar C}, \qquad \quad {\bar f}_2 = \bar C\,\dot e + \dot{\bar C}\,e, \nonumber\\ 
&&{\bar b}_1 = \bar C\,\dot\chi + \dot{\bar C}\,\chi, \quad\qquad {\bar b}_2 = i\,\bar B, \quad\qquad {\bar b}_3 = \bar C\,\dot{\bar C}. 
\end{eqnarray}
In other words, we have already determined the secondary variables (i.e. the coefficients of $\theta$) of the super expansions (62). 
The substitutions of (63) into the {\it chiral} super expansion (62) on the (1,1)-dimensional {\it chiral} super sub-manifold leads to the following
\begin{eqnarray}
&&{\tilde B}^{(ab)}(\tau, \theta) = B(\tau) + \theta\,(\dot B\,\bar C - B\,\dot{\bar C}) \equiv B(\tau) + \theta\,(s_{ab}\,B), \nonumber\\
&&E^{(ab)}(\tau, \theta) = e(\tau) + \theta\,(e\,\dot{\bar C} + \dot e\,\bar C) \equiv e(\tau) + \theta\,(s_{ab}\,e), \nonumber\\
&&K^{(ab)}(\tau, \theta) = \chi(\tau) + \theta\,(\dot{\bar C}\,\chi + \bar C\,\dot\chi) \equiv \chi(\tau) + \theta\,(s_{ab}\,\chi), \nonumber\\
&&F^{(ab)}(\tau, \theta) = C(\tau) + \theta\,(i\,\bar B) \equiv C(\tau) + \theta\,(s_{ab}\,C), \nonumber\\
&&{\bar F}^{(ab)}(\tau, \theta) = \bar C(\tau) + \theta\,(\bar C\,\dot{\bar C}) \equiv \bar C(\tau) + \theta\,(s_{ab}\,\bar C), 
\end{eqnarray}
where the superscript $(ab)$ denotes the {\it chiral} supervariables that have been obtained after the application of the anti-BRST
(i.e. {\it quantum} gauge) invariant restrictions in Eq. (61). It is evident, from the equation (64), that we have already obtained the anti-BRST symmetry 
transformations [cf. Eq. (20)] of the variables $(B, e, \chi, C, \bar C)$ as the coefficients of $\theta$ in the {\it chiral} super expansions (62). We
also observe, in the above chiral super expansions, that there is a mapping: $\partial_\theta \leftrightarrow s_{ab}$ which agrees with
the result of Refs. [10-12].
For the sake of completeness, we perform the step-by-step computations of the mathematical relationships
 obtained   in the equation (63) in our Appendix A.

\section{Symmetry Invariance of the Lagrangians: ACSA}

In this section, we use the results of the previous section to capture the (anti-)BRST invariance of the Lagrangians [cf. Eqs. (26)-(29)] within the 
framework of ACSA. To accomplish this goal, first of all, we generalize the Lagrangians (24) to their counterparts (anti-)chiral {\it super} Lagrangians 
as follows
\begin{eqnarray}  
L_B \rightarrow {\tilde L}_B^{(ac)}(\tau, \bar\theta) &=& {\tilde L}_f^{(ac)}(\tau, \bar\theta) + {\tilde B}^{(b)}(\tau, \bar\theta)
\Big[E^{(b)}(\tau, \bar\theta)\,{\dot E}^{(b)}(\tau, \bar\theta) 
- i\,\big\{2\,{\dot{\bar F}}^{(b)}(\tau, \bar\theta)\,F^{(b)}(\tau, \bar\theta) \nonumber\\ 
&+& {\bar F}^{(b)}(\tau, \bar\theta)\,{\dot F}^{(b)}(\tau, \bar\theta)\big\}\Big] + \frac{1}{2}\,{\tilde B}^{(b)}(\tau, \bar\theta)\,{\tilde B}^{(b)}(\tau, \bar\theta)
\nonumber\\
&-& i\,E^{(b)}(\tau, \bar\theta)\,E^{(b)}(\tau, \bar\theta)\,{\dot{\bar F}}^{(b)}(\tau, \bar\theta)\,{\dot F}^{(b)}(\tau, \bar\theta) \nonumber\\
&-& i\,E^{(b)}(\tau, \bar\theta)\,{\dot E}^{(b)}(\tau, \bar\theta)\,{\dot{\bar F}}^{(b)}(\tau, \bar\theta)\,{F}^{(b)}(\tau, \bar\theta) \nonumber\\
&-& {\dot{\bar F}}^{(b)}(\tau, \bar\theta)\,{\bar F}^{(b)}(\tau, \bar\theta)\,{\dot F}^{(b)}(\tau, \bar\theta)\,F^{(b)}(\tau, \bar\theta),
\end{eqnarray}
\begin{eqnarray}  
L_{\bar B} \rightarrow {\tilde L}_{\bar B}^{(c)}(\tau, \theta) &=& {\tilde L}_f^{(c)}(\tau, \theta) - {\tilde{\bar B}}^{(ab)}(\tau, \theta)\Big[E^{(ab)}
(\tau, \theta)\,
{\dot E}^{(ab)}(\tau, \theta) 
- i\,\big\{2\,{{\bar F}}^{(ab)}(\tau, \theta)\,{\dot F}^{(ab)}(\tau, \theta) \nonumber\\ 
&+& {\dot{\bar F}}^{(ab)}(\tau, \theta)\,{F}^{(ab)}(\tau, \theta)\big \}\Big] + \frac{1}{2}\,{\tilde{\bar B}}^{(ab)}(\tau, \theta)\,{\tilde{\bar B}}^{(ab)}
(\tau, \theta)\nonumber\\
&-& i\,E^{(ab)}(\tau, \theta)\,E^{(ab)}(\tau, \theta)\,{\dot{\bar F}}^{(ab)}(\tau, \theta)\,{\dot F}^{(ab)}(\tau, \theta) \nonumber\\
&-& i\,E^{(ab)}(\tau, \theta)\,{\dot E}^{(ab)}(\tau, \theta)\,{{\bar F}}^{(ab)}(\tau, \theta)\,{\dot F}^{(ab)}(\tau, \theta) \nonumber\\
&-& {\dot{\bar F}}^{(ab)}(\tau, \theta)\,{\bar F}^{(ab)}(\tau, \theta)\,{\dot F}^{(ab)}(\tau, \theta)\,F^{(ab)}(\tau, \theta),
\end{eqnarray}
where the superscripts $(ac)$ and $(c)$, on the r.h.s., denote the {\it anti-chiral} and {\it chiral} versions of the first-order Lagrangian $L_f$ [cf. Eq. (1)]. In other words, we have the following  
\begin{eqnarray}
L_f \rightarrow {\tilde L}_f^{(ac)}(\tau, \bar\theta) &=& P_\mu^{(b)}(\tau, \bar\theta)\,{\dot X}^{\mu{(b)}}(\tau, \bar\theta) - \frac{1}{2}\,E^{(b)}(\tau, \bar\theta)\,
\Big[P_\mu^{(b)}(\tau, \bar\theta)\,P^{\mu{(b)}}(\tau, \bar\theta) - m^2\Big] \nonumber \\
& + & \frac{i}{2}\,\Big[\Psi_\mu^{(b)}(\tau, \bar\theta)\,{\dot\Psi}^{\mu{(b)}}(\tau, \bar\theta) 
- \Psi_5^{(b)}(\tau, \bar\theta)\,{\dot\Psi}_5^{(b)} (\tau, \bar\theta)\Big] \nonumber \\
& + & i\,{\tilde\chi}^{(b)}(\tau, \bar\theta)\,\Big[P_\mu^{(b)}(\tau, \bar\theta)\,\Psi^{\mu{(b)}}(\tau, \bar\theta) 
- m\,\Psi_5^{(b)}(\tau, \bar\theta)\Big],
\end{eqnarray}
\begin{eqnarray}
L_f \rightarrow {\tilde L}_f^{(c)}(\tau, \theta) &=& P_\mu^{(ab)}(\tau, \theta)\,{\dot X}^{\mu{(ab)}}(\tau, \theta) - \frac{1}{2}\,E^{(ab)}(\tau, \theta)\,
\Big[P_\mu^{(ab)}(\tau, \theta)\,P^{\mu{(ab)}}(\tau, \theta) - m^2\Big] \nonumber \\
& + & \frac{i}{2}\,\Big[\Psi_\mu^{(ab)}(\tau, \theta)\,{\dot\Psi}^{\mu{(ab)}}(\tau, \theta) - \Psi_5^{(ab)}(\tau, \theta)\,{\dot\Psi}_5^{(ab)}
(\tau, \theta)\Big] \nonumber \\
& + & i\,{\tilde\chi}^{(ab)}(\tau, \theta)\,\Big[P_\mu^{(ab)}(\tau, \theta)\,\Psi^{\mu{(ab)}}(\tau, \theta) - m\,\Psi_5^{(ab)}(\tau, \theta)\Big],
\end{eqnarray}
where the superscripts $(b)$ and $(ab)$ on the supervariables of ${\tilde L}_f^{(ac)}$ and ${\tilde L}_f^{(c)}$ have been already explained in the previous section. The superscripts $(ac)$ and
$(c)$ on the super Lagrangians, on the l.h.s. of the equations (65) and (66) denote the {\it anti-chiral} and {\it chiral} generalizations of the 
{\it ordinary} Lagrangians (24). Keeping in our mind the mappings: 
$s_b \leftrightarrow \partial_{\bar \theta},\,s_{ab} \leftrightarrow \partial_{\theta}$, we observe the (anti-)BRST invariance of the first-order 
Lagrangian in the language of ACSA to BRST formalism as follows:
\begin{eqnarray}  
\frac{\partial}{\partial\,\bar\theta}\,{\tilde L}_f^{(ac)} = \frac{d}{d\,\tau}\big[C\,L_f\big] \equiv s_b\,L_f, \qquad\quad \frac{\partial}{\partial\,\theta}\,
{\tilde L}_f^{(c)} = \frac{d}{d\,\tau}\big[\bar C\,L_f\big] \equiv s_{ab}\,L_f. 
\end{eqnarray}
In other words, we have accomplished the objective of establishing a precise connection
 between the (anti-)BRST invariance of $L_f$ in the {\it ordinary} space
[cf. Eqs. (3), (27), (26)] and {\it superspace} within the purview  of ACSA.
 The results in (69) will be useful in the proof of the (anti-)BRST
 invariance [cf. Eqs. (27),(26)] of the coupled (but equivalent) Lagrangians 
$L_B$ and $L_{\bar B}$. Geometrically, it is clear from our observation in (69) that super Lagrangians
 ${\tilde L}_f^{(ac, c)}$ are the unique sum of (anti-)chiral supervariables 
[obtained after the (anti-)BRST invariant restrictions] such that
 their translations along $(\theta, \bar\theta)$-directions in the {\it superspace} 
produce the {\it total} derivatives in the {\it ordinary} space.

We now focus on the BRST and anti-BRST invariance of $L_B$ and $L_{\bar B}$ within the purview of ACSA. In the explicit expressions of (65) and (66), 
we substitute the super expansions of (52), (57), (58) and (64) and apply the derivatives $(\partial_{\bar\theta}, \partial_\theta)$ on them due to the
mappings: $\partial_{\bar\theta} \leftrightarrow s_b,\,\, \partial_\theta \leftrightarrow s_{ab}$. It is straightforward to check that we have the
following explicit relationships between the invariances in the {\it superspace} and {\it ordinary} space:
\begin{eqnarray} 
\frac{\partial}{\partial\,\bar\theta}\,{\tilde L}_B^{(ac)} = \frac{d}{d\,\tau}\,\Big[C\,L_f + e^2\,B\,\dot C + e\,\dot e\,B\,C - i\,B\,\bar C\,\dot C\,C + B^2\,C\Big] = s_b\,L_B, \nonumber\\
\frac{\partial}{\partial\,\theta}\,{\tilde L}_{\bar B}^{(c)} = \frac{d}{d\,\tau}\,\Big[\bar C\,L_f - e^2\,\bar B\,\dot{\bar C} - e\,\dot e\,\bar B\,\bar C 
- i\,\bar B\,\dot{\bar C}\,\bar C\,C + {\bar B}^2\,\bar C\Big] = s_{ab}\,L_{\bar B}.
\end{eqnarray}
We would like to emphasize that the super Lagrangian ${\tilde L}_{B}^{(ac)}$ is a unique sum of anti-chiral supervariables (derived after
the applications of the BRST invariant restrictions) such that its translation along $\bar\theta$-direction of the 
$(1, 1)$-dimensional {\it anti-chiral} super sub-manifold leads to a total derivative in the {\it ordinary} space. The {\it latter} is nothing 
but the BRST invariance of the {\it ordinary} Lagrangian $L_B$ [cf. Eq. (26)]. In exactly similar fashion, we can provide a geometrical 
interpretation for the anti-BRST invariance of $L_{\bar B}$ [cf. Eq. (27)] in the terminology of the superspace translational generator
$(\partial_\theta)$ along the $\theta$-direction of the suitably chosen {\it chiral} $(1, 1)$-dimensional super sub-manifold.

At this juncture, we concentrate on the deduction of the CF-type restriction $[B + \bar B + i\,(\bar C\,\dot C - \dot{\bar C}\,C) = 0]$ in the proof of
the {\it equivalence} between the Lagrangians $L_B$ and $L_{\bar B}$ [cf. Eq. (24)] within the ambit of ACSA to BRST formalism. In other words, 
we capture the transformations 
$s_{ab}\,L_B$ [cf. Eq. (29)] and $s_b\,L_{\bar B}$ [cf. Eq. (28)] in the terminology of the ACSA. Towards this central goal 
in our mind, first of all, we generalize the {\it ordinary} Lagrangian $L_B$ to its counterpart {\it chiral} super Lagrangian as
\begin{eqnarray}  
L_B \rightarrow {\tilde L}_B^{(c)}(\tau, \theta) &=& {\tilde L}_f^{(c)}(\tau, \theta) + {\tilde B}^{(ab)}(\tau, \theta)
\Big[E^{(ab)}(\tau, \theta)\,{\dot E}^{(ab)}(\tau, \theta) 
- i\,\{2\,{\dot{\bar F}}^{(ab)}(\tau, \theta)\,F^{(ab)}(\tau, \theta) \nonumber\\ 
&+& {\bar F}^{(ab)}(\tau, \theta)\,{\dot F}^{(ab)}(\tau, \theta)\}\Big] + \frac{1}{2}\,{\tilde B}^{(ab)}(\tau, \theta)\,{\tilde B}^{(ab)}(\tau, \theta)\nonumber\\
&-& i\,E^{(ab)}(\tau, \theta)\,E^{(ab)}(\tau, \theta)\,{\dot{\bar F}}^{(ab)}(\tau, \theta)\,{\dot F}^{(ab)}(\tau, \theta) \nonumber\\
&-& i\,E^{(ab)}(\tau, \theta)\,{\dot E}^{(ab)}(\tau, \theta)\,{\dot{\bar F}}^{(ab)}(\tau, \theta)\,{F}^{(ab)}(\tau, \theta) \nonumber\\
&-& {\dot{\bar F}}^{(ab)}(\tau, \theta)\,{\bar F}^{(ab)}(\tau, \theta)\,{\dot F}^{(ab)}(\tau, \theta)\,F^{(ab)}(\tau, \theta),
\end{eqnarray}   
where the superscripts $(c)$ and $(ab)$ have already been explained in our earlier discussions. It is very interesting to observe that
we have the following:
\begin{eqnarray}
\frac{\partial}{\partial\,\theta}\,{\tilde L}_B^{(c)}(\tau, \theta) &=& \frac{d}{d\,\tau}\,\Big[\bar C\,L_f + e^2\,(i\,\dot{\bar C}\,\bar C\,\dot C 
+ B \,\dot{\bar C}) 
+ e\,\dot e\,(i\,\dot{\bar C}\,\bar C\,C + B\,\bar C) \nonumber\\ 
&+& i\,(2\,B - \bar B)\,\dot{\bar C}\,\bar C\,C + {B}^2\,\bar C\Big] \nonumber\\
&+& \big[B+ \bar B + i\,(\bar C\,\dot C - \dot{\bar C}\,C)\big]\,(i\,\ddot{\bar C}\,\bar C\,C 
- e\,\dot e\,\dot{\bar C} - 2\,B\,\dot{\bar C} + 2\,i\,\dot{\bar C}\,\bar C\,\dot C) \nonumber\\
&-& \frac{d}{d\,\tau}\big[B+ \bar B + i\,(\bar C\,\dot C - \dot{\bar C}\,C)\big]\,(e^2\,\dot{\bar C} + B\,\bar C) \equiv s_{ab}\,L_B. 
\end{eqnarray}
The above equation shows that the Lagrangian $L_B$ respects the anti-BRST symmetry transformations (20) {\it only} when the
CF-type restriction is invoked from {\it outside}. In a subtle manner, we have obtained the CF-type restriction 
$B+ \bar B + i\,(\bar C\,\dot C - \dot{\bar C}\,C) = 0$ within the ambit of ACSA while proving the anti-BRST invariance of the Lagrangian $L_B$. We
now demonstrate the BRST invariance of the Lagrangian $L_{\bar B}$ {\it and} existence of the CF-type restriction: $B + \bar B + i\,(\bar C\,\dot C
- \dot{\bar C}\,C) = 0$ within the purview of theoretical tricks and techniques of ACSA to BRST formalism. Towards these aims in our mind, we
generalize the Lagrangian $L_{\bar B}$ onto $(1, 1)$-dimensional {\it anti-chiral} super submanifold as follows  
\begin{eqnarray}  
L_{\bar B} \rightarrow {\tilde L}_{\bar B}^{(ac)}(\tau, \bar\theta) &=& {\tilde L}_f^{(ac)}(\tau, \bar\theta) - {\tilde{\bar B}}^{(b)}(\tau,
 \bar\theta)\Big[E^{(b)}(\tau, \bar\theta)\,{\dot E}^{(b)}(\tau, \bar\theta) - i\,\big\{2\,{{\bar F}}^{(b)}(\tau, \bar\theta)\,{\dot F}^{(b)}(\tau, \bar\theta)
 \nonumber\\ 
&+& {\dot{\bar F}}^{(b)}(\tau, \bar\theta)\,{F}^{(b)}(\tau, \bar\theta)\big \}\Big] + \frac{1}{2}\,{\tilde{\bar B}}^{(b)}(\tau, \bar\theta)\,
{\tilde{\bar B}}^{(b)}(\tau, \bar\theta)
\nonumber\\
&-& i\,E^{(b)}(\tau, \bar\theta)\,E^{(b)}(\tau, \bar\theta)\,{\dot{\bar F}}^{(b)}(\tau, \bar\theta)\,{\dot F}^{(b)}(\tau, \bar\theta) \nonumber\\
&-& i\,E^{(b)}(\tau, \bar\theta)\,{\dot E}^{(b)}(\tau, \bar\theta)\,{{\bar F}}^{(b)}(\tau, \bar\theta)\,{\dot F}^{(b)}(\tau, \bar\theta) \nonumber\\
&-& {\dot{\bar F}}^{(b)}(\tau, \bar\theta)\,{\bar F}^{(b)}(\tau, \bar\theta)\,{\dot F}^{(b)}(\tau, \bar\theta)\,F^{(b)}(\tau, \bar\theta),
\end{eqnarray}
where the superscripts $(ac)$ and $(b)$ have been already explained in our earlier discussions. Keeping in our mind the mapping: 
$\partial_{\bar\theta} \leftrightarrow s_b$, we observe the following interesting relationship between $s_b\,L_{\bar B}$ and its counterpart
in superspace, namely;  
\begin{eqnarray}
\frac{\partial}{\partial\,\bar\theta}\,{\tilde L}_{\bar B}^{(ac)}(\tau, \bar\theta) &=& \frac{d}{d\,\tau}\,\Big[C\,L_f 
- e^2\,(i\,\dot{\bar C}\,\dot C\, C + \bar B \,\dot C) - e\,\dot e\,(i\,\bar C
\,\dot C\, C + \bar B\,C) \nonumber\\ 
&+& i\,(2\,\bar B - B)\,\bar C\,\dot C\, C + {\bar B}^2\,C\Big] \nonumber\\
&+& \big[B+ \bar B + i\,(\bar C\,\dot C - \dot{\bar C}\,C)\big]\,\big [i\,\bar C\,\ddot C\,C + e\,\dot e\,\dot C - 2\,\bar B\,\dot C + 2\,i\,
\dot{\bar C}\,\dot C\, C \big ] \nonumber\\
&+& \frac{d}{d\,\tau}\big[B+ \bar B + i\,(\bar C\,\dot C - \dot{\bar C}\,C)\big]\,(e^2\,\dot C - \bar B\,C) \equiv s_b\,L_{\bar B}, 
\end{eqnarray}
where the r.h.s. is nothing but the operation of $s_b$ on the Lagrangian $L_{\bar B}$
 [cf. Eq. (28)]. In other words, we have established an 
intimate relationship between the BRST symmetry transformation on $L_{\bar B}$ {\it and}
 operation of the translational generator $\partial_{\bar\theta}$
on the {\it anti-chiral} super Lagrangian ${\tilde L}_{\bar B}^{(ac)}(\tau, \bar\theta)$
 [defined on the $(1, 1)$-dimensional (anti-)chiral super
sub-manifold]. A careful and close look on the r.h.s. of (74) demonstrates that we have 
derived the CF-type restriction: $B+ \bar B + i\,(\bar C\,\dot C -
 \dot{\bar C}\,C) = 0$ while proving the BRST invariance of $L_{\bar B}$ within the purview of ACSA.

We end this section with the following concluding remarks. First, we have captured the (anti-)BRST invariance [cf. Eq. (69)] of the first-order Lagrangian 
$L_f$ within the ambit of ACSA. Second, we have been able to express the (anti-)BRST invariance [cf. Eqs. (26), (27)] of the
Lagrangians $L_B$ and $L_{\bar B}$ of our theory within the purview of ACSA  [cf. Eq. (70)]. Third, we have been able to demonstrate that our
observations in the equations (28) and (29) can {\it also} be expressed in superspace [cf. Eqs. (72), (74)] within the ambit of ACSA. Finally, we have derived the CF-type restriction: $B+ \bar B + i\,(\bar C\,\dot C - \dot{\bar C}\,C) = 0$, in a subtle manner,
by expressing the transformations $s_b\,L_{\bar B}$ and $s_{ab}\,L_B$ in the language of ACSA  [cf. Eqs. (72), (74) for details].
In other words, in the ordinary space, whatever we have seen in the proof of the absolute 
anticommutativity of (anti-)BRST symmetry transformations
[cf. Eq. (22)], the {\it same} restriction appears when we discuss $s_b\,L_{\bar B}$ and 
$s_{ab}\,L_B$ in the {\it superspace} by using 
the formal theoretical techniques of ACSA.

\section{Off-Shell Nilpotency and Absolute Anticommutativity of the Conserved (Anti-)BRST Charges}

In this section, we prove the off-shell nilpotency $[Q_{(\bar B)B}^{2} = 0]$ and absolute anticommutativity $\big[\{Q_B, Q_{\bar B} \} = 0\big]$ 
of the (anti-)BRST charges $Q_{(\bar B)B}$ which have already been derived in our Sec. 3 where $J_B = Q_B$ and $J_{\bar B}= Q_{\bar B}$ [cf. Eqs. (31), (32)]. In
 sub-section 7.1, we discuss the above properties of the (anti-)BRST charges $[Q_{(\bar B)B}]$ in the {\it ordinary} space. Our sub-section 7.2 contains the
 theoretical material related with the techniques of capturing the nilpotency and anticommutativity properties of the above charges within the ambit of ACSA. 
It is quite interesting to state that
 the CF-type restriction appears
when we prove the absolute anticommutativity property of the charges in the {\it ordinary} space as well as 
in the {\it superspace} (within the purview of ACSA).
We describe explicitly the computation of $s_b L_{\bar B}$ in our Appendix B where the algebra is a bit more involved
 and, ultimately, we demonstrate the existence of the CF-type restriction in the {\it ordinary} space [cf. Eq. (28)] which has {\it also} appeared in Eq. (74)
within the ambit of ACSA.

\subsection{Off-Shell Nilpotency and Absolute Anticommutativity Properties: Ordinary Space}

In this sub-section, we primarily exploit the theoretical potential of the well-known relationship between the continuous symmetry 
transformations and their generators. In other words, we can prove the off-shell nilpotency $(Q_B^2 = Q_{\bar B}^2 = 0)$ of the (anti-)BRST 
charges $Q_{(\bar B)B}$ 
in the {\it ordinary}
space by using the standard relationship between the infinitesimal continuous (anti-)BRST transformations $(s_{(a)b})$ and their generators
($Q_{(\bar B)B}$) as:
\begin{eqnarray}
s_b\,Q_B = -\,i\{Q_B, Q_B\} = 0 \quad \implies \quad Q_B^2 = 0, \nonumber\\
s_{ab}\,Q_{\bar B} = -\,i\{Q_{\bar B}, Q_{\bar B}\} = 0 \quad \implies \quad Q_{\bar B}^2 = 0.
\end{eqnarray}
The above proofs of the off-shell nilpotency of the conserved charges are nothing but the reflection of the off-shell nilpotency 
$(s_{(a)b}^2 = 0)$ of the (anti-)BRST symmetry transformations (20) and (21) in the {\it ordinary} space.
It would be worthwhile to point out the fact that, in the computation of the l.h.s. of (75), we have {\it directly}
applied the (anti-)BRST symmetry transformations (20) and (21) on the appropriate form of the conserved  (i.e. $\dot Q_{(\bar B)B} = 0$) 
(anti-)BRST charges [cf. Eqs. (31), (32)]. In other words, the straightforward application of $s_b$ on $Q_B$
gives us a {\it zero} result. Same is the situation (i.e. $s_{ab} Q_{\bar B} = 0$) when we apply the anti-BRST symmetry transformations $s_{ab}$
on the anti-BRST charge $Q_{\bar B}$. Hence, the Noether conserved charges [cf. Eqs. (31), (32)] are 
off-shell nilpotent of order two (i.e. $Q_{(\bar B)B}^2 = 0$) in the {\it ordinary} space due to the key relationship that is given in (75).

The above explicit proof of the off-shell nilpotency of the (anti-)BRST charges ensures that they should be able to be written as {\it an exact}
quantity w.r.t. the off-shell nilpotent $[s_{(a)b}^2 = 0]$ (anti-)BRST symmetry transformations $[s_{(a)b}]$. Towards this goal in mind, 
we use the following EL-EOMs (derived from the Lagrangians $L_B$ and $L_{\bar B}$), namely:
\begin{eqnarray}
&& p_\mu\,\psi^\mu = m\,\psi_5,  \qquad
 B = - e\,\dot e + 2\,i\,\dot{\bar C}\,C + i\,\bar C\,\dot C, \nonumber\\
&& e\,\dot B + i\,e\,\dot{\bar C}\,\dot C - i\,e\,\ddot{\bar C}\,C + \frac{1}{2}\,(p^2 - m^2) = 0, \nonumber \\
&& B\,\dot C + 2\,\dot B\,C + 3\,e\,\dot e\,\dot C + e^2\,\ddot C + {\dot e}^2\,C + e\,\ddot e\,C - i\,\bar C\,\ddot C\,C
- 2\,i\,\dot{\bar C}\,\dot C\,C = 0,\nonumber\\
&& \bar B = e\,\dot e - 2\,i\,\bar C\,\dot C - i\,\dot{\bar C}\,C,  \quad e\,\dot{\bar B} - i\,e\,\dot{\bar C}\,\dot C + i\,e\,{\bar C}\,\ddot C - 
\frac{1}{2}\,(p^2 - m^2) = 0, \nonumber \\
&& \bar B\,\dot{\bar C} + 2\,\dot{\bar B}\,\bar C - 3\,e\,\dot e\,\dot{\bar C} - e^2\,\ddot{\bar C} - {\dot e}^2\,\bar C 
- e\,\ddot e\,\bar C - i\,\ddot{\bar C}\,\bar C\,C
- 2\,i\,\dot{\bar C}\,\bar C\,\dot C = 0,
\end{eqnarray}
to get rid of the constraints $(p^2 - m^2) \approx 0$ and $(p_\mu\,\psi^\mu - m\,\psi_5) \approx 0$ 
from the expressions for the (anti-)BRST charges $Q_{(\bar B)B}$ [cf. Eqs. (31), (32)] to recast them as
\begin{eqnarray}
&&Q_{\bar B}^{(1)} = e^2\,\Big[\dot{\bar B}\,\bar C - \bar B\,\dot{\bar C} + i\,(\ddot{\bar C}\,\bar C\,C +\dot{\bar C}\,\bar C\,\dot C)\Big]  +
2\,i\,e\,\dot e\,\dot{\bar C}\,\bar C\,C, \nonumber\\
&&Q_B^{(1)} = e^2\,\Big[B\,\dot C - \dot B\,C - i\,(\bar C\,\ddot C\,C + \dot{\bar C}\,\dot C\,C)\Big] - 2\,i\,e\,\dot e\,\bar C\,\dot C\,C.
\end{eqnarray}
We have discussed different forms of the (anti-)BRST charges in our Appendix 
C where the emphasis is laid on the derivation of  the expressions for the (anti)BRST charges [cf. Eq. (77)].
The above expressions of the conserved (anti-)BRST charges can be mathematically expressed in the
following {\it exact} forms w.r.t. the off-shell nilpotent $[s_{(a)b}^2 = 0]$ (anti-)BRST symmetry
transformations $[s_{(a)b}]$, namely;
\begin{eqnarray}
Q_{\bar B}^{(1)} = s_{ab}\,[i\,e^2\,({\bar C}\,\dot C - \dot{\bar C}\,C)], \qquad Q_B^{(1)} = s_b\,[i\,e^2\,(\dot{\bar C}\,C - \bar C\,\dot C)].
\end{eqnarray}
Now it is straightforward to note that $s_{ab}\,Q_{\bar B}^{(1)} = 0$ and 
$s_{b}\,Q_{B}^{(1)} = 0$ due to the off-shell nilpotency [$s_{(a)b}^{2} = 0$] of the
(anti-)BRST symmetry transformations $[s_{(a)b}]$. Thus, we conclude, from our observations 
in Eq. (78), that the nilpotency of the (anti-)BRST transformations [$s_{(a)b}$] is deeply 
connected with the nilpotency of their generators (anti-)BRST charges [$Q_{(\bar B)B}$] which becomes completely 
transparent from the {\it direct} observations of the following computations:
\begin{eqnarray}
 && \big[Q_{\bar B}^{(1)}\big]^2 = 0 \quad \Longleftrightarrow  \quad s_{ab}\,Q_{\bar B}^{(1)}
 = -\,i\,\big\{Q_{\bar B}^{(1)}, Q_{\bar B}^{(1)}\big\} = 0  
\quad \Longleftrightarrow  \quad s_{ab}^2 = 0, \nonumber\\
 && \big[Q_{B}^{(1)}\big]^2 = 0  \quad \Longleftrightarrow  \quad s_{b}\,Q_{B}^{(1)} = -\,i\,\big\{Q_{B}^{(1)}, Q_{B}^{(1)}\big\} = 0 
\quad\; \Longleftrightarrow  \quad s_{b}^2 = 0.
\end{eqnarray}
The above equation completes our discussion on the proof of the off-shell nilpotency 
of the conserved (anti-)BRST charges in the {\it ordinary} space.

Now we dwell on the proof of the absolute anticommutativity (i.e. $Q_{B}^{(1)}\,Q_{\bar B}^{(1)} 
+  Q_{\bar B}^{(1)}\,Q_{B}^{(1)} = 0$) of the conserved off-shell
nilpotent (anti-)BRST charges [$Q_{(\bar B)B}^{(1)}$]. Towards this central objective in
mind, first of all, we assume the sanctity and validity of the CF-type
restriction; $B + \bar B + i\,(\bar C\, \dot C - \dot{\bar C}\,C) = 0$, {\it right from the beginning}. As a
result, we can express the (anti-) BRST charges $Q^{(1)}_{(\bar B)B}$, in the alternative forms, as follows:
\begin{eqnarray}
Q_{\bar B}^{(1)} \;\rightarrow \; Q_{\bar B}^{(2)} = e^2\,(B\,\dot{\bar C} - \dot B\,\bar C + 2\,i\,\dot{\bar C}\,\bar C\,\dot C) + 2\,i\,e\,\dot e\,\dot{\bar C}\,\bar C\,C, \nonumber\\
Q_{ B}^{(1)} \; \rightarrow \; Q_B^{(2)} = e^2\,(\dot{\bar B}\,C - \bar B\,\dot C - 2\,i\,\dot{\bar C}\,\dot C\,C) - 2\,i\,e\,\dot e\,\bar C\,\dot C\,C.
\end{eqnarray}
We point out that it is because of the use of the CF-type restriction [$B + \bar B + i\,(\bar C\,\dot C - \dot{\bar C}\,C) = 0$ ] that 
we have been able to express $ Q_{\bar B}^{(2)}$ in terms of the Nakanishi-Lautrap 
auxiliary variable $B(\tau)$ and $ Q_{B}^{(2)}$ in the language of other
Nakanishi-Lautrap type auxiliary variable $\bar B(\tau)$. At this crucial stage, 
we observe the following interesting relationships:
\begin{eqnarray}
  Q_{B}^{(2)} = s_{ab}\,\Big[-i\,\,e^2\,\dot{C}\,C\Big], \qquad\qquad Q_{\bar B}^{(2)} = s_b\,\Big[i\,e^2\,\dot{\bar C}\,\bar C\Big], 
\end{eqnarray}
In other words, we have been able to express the anti-BRST charge [$ Q_{\bar B}^{(2)}$]
 as the BRST {\it exact} quantity. On the other hand, we have been 
able to write the BRST charge [$ Q_{B}^{(2)}$] as an {\it exact} quantity w.r.t. 
.the nilpotent anti-BRST transformation $s_{ab}$. A close and careful observation of (81) leads
 to the following (due to the well-known relationship between the continuous (anti-)BRST 
symmetry transformations $[s_{(a)b}]$ and their generators as
conserved (anti-)BRST charges $[Q_{(\bar B)B}^{(2)}]$), namely;
\begin{eqnarray}
s_{ab}\,Q_{B}^{(2)} &=& -\,i\,\big\{Q_{B}^{(2)}, Q_{\bar B}^{(2)}\big\} = 0 \quad \Longleftrightarrow  \quad s_{ab}^2 = 0,  \nonumber\\
s_{b}\,Q_{\bar B}^{(2)} &=& -\,i\,\big\{Q_{\bar B}^{(2)}, Q_{B}^{(2)}\big\} = 0 \quad \Longleftrightarrow  \quad s_b^2 = 0.
\end{eqnarray}
As a result, we observe that the absolute anticommutativity of the (anti-)BRST charges $[Q_{(\bar B)B}^{(2)}]$ is related  to the
 nilpotency  $[s_{(a)b}^2] = 0$ of the (anti-)BRST symmetries.

We would like to lay stress  on the key results  that have been seen in Eq. (82). It is very interesting 
(due to the validity of the CF-type restriction on our theory) to pinpoint that (i) the anticommutativity of the BRST 
charge $Q_B^{(2)}$ {\it with} the anti-BRST charge $Q_{\bar B}^{(2)}$ is intimately connected with the
nilpotency $(s_b^{2} = 0)$ of the BRST transformations $(s_b)$, and (ii) the anticommutativity property 
of the anti-BRST charge $Q_{\bar B}^{(2)}$ {\it with} the BRST charge $Q_B^{(2)}$ owes its origin to the nilpotency
$(s_{ab}^2 = 0)$ of the anti-BRST transformations $(s_{ab})$. We conclude this sub-section with the following 
remarks. First, we have shown that the nilpotency of the (anti-)BRST charges $[Q_{(\bar B)B}]$ is deeply related
with the nilpotency of the (anti-)BRST transformations $[s_{(a)b}]$. Second, we have been able to
express the {\it modified} form of the BRST charge $[Q_B^{(1)}]$ and anti-BRST charge $Q_{\bar B}^{(1)}$ as the {\it exact}
expressions w.r.t. the BRST transformations $(s_b)$ and anti-BRST transformations $(s_{ab})$ [cf. Eq. (78)], respectively. Third,
it is due to the {\it existence} of the CF-type restriction on our theory that we have been able to express {\it another}
modified form of the BRST charge $Q_B^{(2)}$ as {\it an} exact expression w.r.t. the anti-BRST transformations
$(s_{ab})$ {\it and} the anti-BRST charge $Q_{\bar B}^{(2)}$ in the BRST-{\it exact} form. This exercise has enabled us to
prove the absolute anticommutativity [i.e. $\{Q_B^{(2)}, Q_{\bar B}^{(2)}\} = 0$] of the nilpotent (anti-)BRST charges $Q_{(\bar B)B}^{(2)}$.
Finally, the proof of the absolute anticommutativity property [cf. Eq. (82)] {\it crucially} depends on the existence of
the CF-type restriction. Thus, in a subtle manner, we have {\it derived} and {\it corroborated} the sanctity of the 
existence of the CF-type restriction $B + \bar B + i\,(\bar C\,\dot C - \dot{\bar C}\,C) = 0$ on our theory.
We have provided an alternative proof for the appearance of the CF-type restriction (on our 1D reparameterization 
invariant SUSY theory) in our Appendix D. This completes our discussions on the  absolute anticommutativity  
property of the conserved (anti-)BRST charges (in the {\it ordinary} Minkowski  space).

\subsection{Off-Shell Nilpotency and Absolute Anticommutativity Properties: ACSA to BRST Formalism in Superspace}

In this sub-section, we capture the properties of the nilpotency (i.e. fermionic nature)
and absolute anticommutativity (i.e. linear independence) 
of the (anti-)BRST charges within the purview of ACSA where the superspace
consideration on the (1, 1)-dimensional (anti-) chiral super submanifolds 
has been taken into account. First of all, we focus on the off-shell nilpotency [$Q_{(\bar B)B}^{2} = 0$] of the (anti-)BRST charges 
[$Q_{(\bar B)B}$]. In this context, keeping in our knowledge the mappings:
$\partial_{\theta}  \leftrightarrow s_{ab}$, $\partial_{\bar \theta}
\leftrightarrow s_b$, it can be readily seen that the expressions for the
(anti-)BRST charges that have been quoted in Eq. (78) can be translated into the superspace as follows
\begin{eqnarray}
&&Q_{\bar B}^{(1)}  =  \frac{\partial}{\partial \,\theta}\,\Big[i\,E^{(ab)}(\tau, \theta)\, E^{(ab)}(\tau, \theta)\,\big \{ \bar F^{(ab)} (\tau, \theta)\,\dot F^{(ab)} (\tau, \theta) - \dot{\bar F}^{(ab)} (\tau, \theta)\,F^{(ab)} (\tau, \theta)  \big \}\Big] \nonumber\\
 &&\equiv  \int d\,\theta \, \Big[i\,E^{(ab)}(\tau, \theta)\, E^{(ab)}(\tau, \theta)\,\big \{ \bar F^{(ab)} (\tau, \theta)\,\dot F^{(ab)} (\tau, \theta) - \dot{\bar F}^{(ab)} (\tau, \theta)\,F^{(ab)} (\tau, \theta)  \big \}\Big],
\end{eqnarray}
\begin{eqnarray}
&&Q_{B}^{(1)}  =  \frac{\partial}{\partial \,\bar \theta}\,\Big[i\,E^{(b)}(\tau, \bar \theta)\, E^{(b)}(\tau, \bar \theta)\,\big \{ \bar F^{(b)} (\tau, \bar \theta)\,\dot F^{(b)} (\tau, \bar \theta) - \dot{\bar F}^{(b)} (\tau, \bar \theta)\,F^{(b)} (\tau, \bar \theta)  \big \}\Big] \nonumber\\
&&\equiv  \int d\,\bar \theta \, \Big[i\,E^{(b)}(\tau, \bar \theta)\, E^{(b)}(\tau, \bar \theta)\,\big \{ \bar F^{(b)} (\tau, \bar \theta)\,\dot F^{(b)} (\tau, \bar \theta) - \dot{\bar F}^{(b)} (\tau, \bar \theta)\,F^{(b)} (\tau, \bar \theta)  \big \}\Big],
\end{eqnarray}
where the supervariables with the superscripts $(ab)$ and $(b)$ have been obtained in Eqs. (64) and (57), respectively. At this stage, the off-shell 
nilpotency $[(Q_{(\bar B)B}^{(1)})^2 = 0]$ of the conserved (anti-)BRST charges $[Q_{(\bar B)B}^{(1)}]$ can be written in the superspace (by using 
the theoretical techniques and tricks of ACSA to BRST formalism) as:
\begin{eqnarray}
\partial_\theta\,Q_{\bar B}^{(1)} = 0 \quad \Longleftrightarrow \quad \partial_\theta^2 = 0, \qquad\qquad  
\partial_{\bar\theta}\,Q_{B}^{(1)} = 0
 \quad \Longleftrightarrow \quad \partial_{\bar\theta}^2 = 0. 
\end{eqnarray}
Thus, we conclude that the off-shell nilpotency of the anti-BRST charge $(Q_{\bar B}^{(1)})$ is deeply related to
the nilpotency $(\partial_\theta^2 = 0)$ of the translational generator $(\partial_\theta)$ along the $\theta$-direction of the $(1, 1)$-
dimensional {\it chiral} super sub-manifold of the {\it general} $(1, 2)$-dimensional supermanifold. Similar type of 
comments can be made in the context of the off-shell nilpotency of the BRST charge $Q_B^{(1)}$ and its intimate
relationship with the nilpotency $(\partial_{\bar\theta}^2 = 0)$ of the translational generator $(\partial_{\bar\theta})$ 
on the {\it anti-chiral} super sub-manifold.

We concentrate now on capturing the absolute anticommutativity of the (anti-)BRST charges within the purview of ACSA
where the superspace of the $(1, 1)$-dimensional (anti-) chiral super sub-manifolds are taken into consideration. Towards
this central goal in our mind, we express the modified forms of the (anti-)BRST charges $Q_{(\bar B)B}^{(2)}$ of Eq. (81) 
in the following mathematical expression within the framework of ACSA, namely;
\begin{eqnarray}
Q_{\bar B}^{(2)} &=& \frac{\partial}{\partial\,\bar\theta}\Big[i\,E^{(b)}(\tau, \bar\theta)\,E^{(b)}(\tau, \bar\theta)\,
\dot{\bar F}^{(b)}(\tau, \bar\theta)\,{\bar F}^{(b)}(\tau, \bar\theta)\Big]\nonumber\\
&\equiv& \int d\,\bar\theta \,\Big[i\,E^{(b)}(\tau, \bar\theta)\,E^{(b)}(\tau, \bar\theta)\,
\dot{\bar F}^{(b)}(\tau, \bar\theta)\,{\bar F}^{(b)}(\tau, \bar\theta)\Big], 
\end{eqnarray}  
\begin{eqnarray}
Q_{B}^{(2)} &=& \frac{\partial}{\partial\,\theta}\Big[-\,i\,E^{(ab)}(\tau, \theta)\,E^{(ab)}(\tau, \theta)\,
\dot{F}^{(ab)}(\tau, \theta)\,{F}^{(ab)}(\tau, \theta)\Big]\nonumber\\
&\equiv& \int d\,\theta \,\Big[-\,i\,E^{(ab)}(\tau, \theta)\,E^{(ab)}(\tau, \theta)\,
\dot{F}^{(ab)}(\tau, \theta)\,{F}^{(ab)}(\tau, \theta)\Big], 
\end{eqnarray}  
where the supervariables with the superscripts $(ab)$ and $(b)$ have been quoted in Eqs. (64) and (57), respectively. 
At this crucial juncture, we note the following:
\begin{eqnarray}
\partial_{\bar\theta}\,Q_{\bar B}^{(2)} = 0 \quad \Longleftrightarrow \quad \partial_{\bar\theta}^2 = 0, \qquad\qquad  
\partial_{\theta}\,Q_{B}^{(2)} = 0
 \quad \Longleftrightarrow \quad \partial_{\theta}^2 = 0. 
\end{eqnarray}
The above relations are nothing but the explicit proof of the absolute  anticommutativity relations  of the conserved  (anti-)BRST charges
$Q_{(\bar B)B}^{(2)}$ (within the ambit of ACSA).

We wrap-up this sub-section with the following comments. First, the off-shell nilpotency $[Q_{(\bar B)B}^2 = 0]$ of the
(anti-)BRST charges $Q_{(\bar B)B}$ is intimately connected with the nilpotency $(\partial_\theta^2 = 0, \partial_{\bar\theta
}^2 = 0)$ of the translational generators $(\partial_\theta, \partial_{\bar\theta})$ along the $(\theta, \bar\theta)$-directions
of the $(1, 1)$-dimensional {\it chiral} and {\it anti-chiral} super sub-manifolds.
 Second, in the {\it ordinary} space, the above statements
of the off-shell nilpotency are captured in the equations (75) and (79). Third, the absolute anticommutativity of the 
BRST charge {\it with} the anti-BRST charge is related to the nilpotency $(\partial_\theta^2 = 0)$ of the translational
generator $(\partial_\theta)$ along the $\theta$-direction of the {\it chiral}
 super submanifold. The absolute anticommutativity
of the anti-BRST charge {\it with} the BRST charge, on the other hand,  is connected 
with the nilpotency $(\partial_{\bar\theta}^2 = 0)$ of the
translational generator $(\partial_{\bar\theta})$ along the $\bar\theta$-direction of the {\it anti-chiral} super submanifold. Fourth, 
the above statements have been corroborated in the {\it ordinary} space by the
 equation (82) where the off-shell nilpotency $[s_{(a)b}^2 = 0]$ of the
(anti-)BRST transformations $[s_{(a)b}]$ and the anticommutativity $[\{Q_B^{(2)}, Q_{\bar B}^{(2)}\} = 0]$ of the (anti-)BRST 
charges $[Q_{(\bar B)B}^{(2)}]$ are found to be inter-connected in an intimate and beautiful  manner.

\section{Conclusions}

The USFA to BRST formalism (see, e.g. [10-12]) is useful in the context of the gauge theories where the spacetime coordinates
 do {\it not} change. Thus,
it was a challenge to include the diffeomorphism (i.e. the general spacetime transformations) within the framework of Bonora-Tonin (BT) 
superfield approach to BRST formalism (see, e.g. [10-12]). This was achieved by Bonora in Ref. [22] which has been christened by us as the MBTSA 
where the generalization of the 1D diffeomorphism [i.e. $ \tau \rightarrow  \tau \,' = f\,(\tau) \,\equiv \,\tau - \varepsilon \,(\tau) $]
to the (1, 2)-dimensional supermanifold [cf. Eq. (35)] has played an important role in the derivation of the (anti-)BRST symmetry 
transformations [cf. Eq. (50)] for the target space variables ($ x_\mu,\, p_\mu,\, \psi_\mu,\, \psi_5$). In addition, this approach 
has enabled us to deduce the (anti-)BRST invariant CF-type restriction: $B + \bar B + i\,(\bar C\,\dot C - \dot{\bar C}\,C) = 0$ 
that is responsible for the absolute anticommutativity of the (anti-)BRST symmetry transformations 
[cf. Eqs. (20), (21)] and existence of the coupled
(but equivalent) Lagrangians (24) for our theory.

We have taken into account the standard (anti-)BRST symmetry transformations ($s_b\,\bar C = i\, B, \, s_{ab}\, C = i\, \bar B$) for 
the (anti-)ghost variables $(\bar C)C$ which have, in a subtle manner, forced us to consider the {\it (anti-)chiral} super expansions [cf. Eq. (47)].
This has provided us the clue to adopt the ACSA to BRST formalism for the deduction of the proper (anti-)BRST transformations for the {\it rest} 
of the variables of our theory (cf. Sec. 5). Within the purview of ACSA, we have derived the CF-type restriction when
we have proven the {\it equivalence} of the coupled (but equivalent) Lagrangians (cf. Sec. 6). Furthermore, it is the validity
of the CF-type restriction: $B + \bar B + i\,(\bar C\,\dot C - \dot{\bar C}\,C) = 0$ that has enabled us to write 
(i) the BRST charge as an {\it exact} quantity w.r.t. the nilpotent anti-BRST transformation, {\it and} (ii) the
anti-BRST charge as an {\it exact} expression w.r.t. the nilpotent BRST transformation. These observations have been
{\it responsible} for the proof of the absolute anticommutativity of the (anti-)BRST charges (cf. Sec. 7). In other
words, it is the {\it proof} of the anticommutativity of the conserved and nilpotent charges
$[Q_{(\bar B)B}^{(2)}]$ which leads to the existence of the CF-type restriction on our SUSY theory (cf. Sec. 7).

We would like to emphasize that the {\it observation} of the absolute anticommutativity property,
in the context of the conserved (anti-)BRST charges, is a novel 
 observation because of the fact that {\it only} the (anti-)chiral super expansions have been considered
 within the ambit of ACSA. This observation of the absolute anticommutativity
property is obvious when one takes into account the {\it full} super expansions of the supervariables {\it along all} the possible
Grassmannian directions of the suitably chosen supermanifold on which the {\it ordinary} theory is generalized.
Furthermore, the appearance of the CF-type restriction in the computations of $s_b\,L_{\bar B}$ and $s_{ab}\,L_B$
[cf. Eqs. (28), (29)] in the {\it ordinary} space {\it and} its analogue in the {\it superspace} are very interesting observations in our present endeavor (cf. Sec. 6). The other observation that merits a clear and special mention
is the {\it universality} of the CF-type restriction: $B + \bar B + i(\bar C\,\dot C - \dot{\bar C}\,C) = 0$ in the context
of the reparameterization (i.e. 1D diffeomorphism) invariant non-SUSY theory of a scalar relativistic particle as well as a non-relativistic particle [24, 23] and our present
SUSY system of a spinning relativistic particle (where the nilpotent SUSY transformations exist between the specific  set of
 bosonic and fermionic variables of our SUSY theory).

It is worthwhile to mention that, for the D-dimensional diffeomorphism invariant theory [22, 26] where the infinitesimal 
diffeomorphism symmetry transformation is: $x_\mu \rightarrow x_\mu' = x_\mu - \epsilon _\mu(x)$ (with $\mu = 0, 1, 2\,...\,D-1$), the {\it general}
 form of the CF-type restriction has been obtained as: $B_\mu + \bar B_{\mu} + i\,(\bar C^{\rho}\,\partial_{\rho}\,C_{\mu} +
 C^{\rho}\,\partial_{\rho}\,\bar C_{\mu}) = 0$ where the (anti-)ghost fields $(\bar C_\mu)C_\mu$ correspond to the 
infinitesimal transformation parameter $\epsilon _\mu(x)$ in the general coordinate transformation: $x _\mu' = x_\mu - \epsilon_\mu (x)$ {\it and} 
the Nakanishi-Lautrup fields $(\bar B_\mu)B_\mu$ appear in the (anti-)BRST symmetry transformations: 
$s_b\, \bar C_\mu = i\,B_\mu$, $s_{ab}\, C_\mu = i\,\bar B_\mu$.
It is straightforward to note that the CF-type restriction: $B + \bar B + i(\bar C\,\dot C - \dot{\bar C}\,C) = 0$ is the limiting case of the
above {\it general} D-dimensional CF-type restriction in the case of the BRST approach to D-dimensional diffeomorphism invariant theory [22, 26]. Thus, our
theoretical treatments of the reparameterization (i.e. 1D diffeomorphism) invariant theories of the {\it scalar} and {\it spinning} 
relativistic particles
are correct.

One of the highlights of ACSA to BRST formalism is the observation that it distinguishes between the suitably chosen $(1, 1)$-dimensional
{\it chiral} and {\it anti-chiral} super sub-manifolds in the proof of the absolute anticommutativity of the conserved 
(anti-)BRST charges. For instance, we note that the anticommutativity of the BRST charge $[Q_B^{(2)}]$ {\it with} the anti-BRST
charge $[Q_{\bar B}^{(2)}]$ is connected with the nilpotency $(\partial_{\theta}^2 = 0)$ of the translational generator 
$(\partial_{\theta})$ along the $\theta$-direction of the {\it chiral} super submanifold [cf. Eq. (88)]. On the other hand, the anticommutativity
of the anti-BRST charge $[Q_{\bar B}^{(2)}]$ {\it with} the BRST charge $[Q_B^{(2)}]$ crucially depends on the nilpotency
$(\partial^2_{\bar\theta} = 0)$ of the translational generator $(\partial_{\bar\theta})$ [cf. Eq. (88)] along the $\bar\theta$-direction of the
{\it anti-chiral} super sub-manifold (cf. Sec. 7 for details). This observation is a reflection of our discussion on the absolute
anticommutativity property of the (anti-)BRST charges in the {\it ordinary} space (cf. Sec. 7) where the off-shell nilpotency of the (anti-)BRST transformations
 [cf. Eq. (82)] play a decisive role.

We plan to extend our present study to the physical (3+1)-dimensional (4D) theories of the gravitation and higher dimensional 
(super)string theories where there is existence of the diffeomorphism invariance. In other words, we plan  to 
apply the ideas of MBTSA and ACSA {\it together} to find out the (anti-)BRST symmetries 
of the above mentioned theories. The mathematical elegance, rigor 
and beauty of the MBTSA [22, 26] should find more applications to some physical 
systems of interest in theoretical high energy physics. We envisage 
to take up these challenges in our future investigations.
Before we end this section, it is worthwhile to point out that in our earlier works (see, e.g. [27, 28]), we have applied
the techniques and tricks of ACSA to obtain the nilpotent symmetries of the 
${\cal N} = 2$ SUSY quantum mechanical models of interest. However, we have found that
the conserved charges are {\it not} absolutely anticommuting. Thus, our observation 
of the absolute anticommutativity property in the context of 
(anti-)BRST charges is {\it novel} and interesting.\\

\begin{center}
{\bf Appendix A: On the Step-by-Step Computation of the Secondary Variables for the Off-shell Nilpotent Anti-BRST Transformations}\\
\end{center}
\vskip 0.5cm
In this Appendix, we concentrate on the clear-cut derivation of the secondary variables [cf. Eq. (63)] in terms of the basic and 
auxiliary variables of the Lagrangians (24). For this purpose, we invoke the basic principle of ACSA which states that the anti-BRST 
invariant expressions [cf. Eq. (59)] should not depend on $\theta$ (i.e. Grassmannian variable) when these quantities are promoted
onto the $(1, 1)$-dimensional {\it chiral} version of super sub-manifold. First of all, 
we consider $s_{ab}\,(\bar C\,{\dot x}_\mu) = 0$ which leads to the following  restriction:
 \[
\bar F(\tau, \theta)\,{\dot X}_\mu^{(hc)}(\tau, \theta) = \bar C(\tau)\,{\dot x}_\mu(\tau).   \eqno(A.1)   
\] 
At this stage, we substitute the super expansions from (62) and (58) which leads to the precise determination of the secondary variable
${\bar b}_3 = \bar C\,\dot{\bar C}$. As a consequence, we have now the super expansion of 
the chiral supervariable $\bar F(\tau, \theta)$ as
\[
{\bar F}^{(ab)}(\tau, \theta) = \bar C(\tau) + \theta\,(\bar C\,\dot{\bar C}) \equiv  \bar C(\tau) + \theta\,(s_{ab}\,\bar C), \eqno (A.2) 
\] 
where the superscript $(ab)$ denotes the {\it chiral} super expansion of $\bar F(\tau, \theta)$ after the application of the anti-BRST 
restriction (A.1). We find that the coefficient of $\theta$ is the anti-BRST symmetry transformation of $\bar C$ [cf. Eq. (20)].
In other words, we find that $\partial_{\theta} \, {\bar F}^{(ab)}(\tau, \theta) =  s_{ab} \bar C$ which agrees 
with the mapping: $\partial_\theta \leftrightarrow s_{ab}$ of  Refs.  [10-12].

At this juncture, we take up the anti-BRST invariant quantities $\big(i.e. s_{ab}\,[ \,d/d\,\tau\,(\bar C\, e)] = 0,\,
s_{ab}\,[\,d/d\,\tau\,(\bar C\, \chi)] = 0\big)$ which leads to the following restrictions:
\[
\frac{d}{d\,\tau}\,\Big[ \bar F^{(ab)}(\tau, \theta)\, E(\tau, \theta) \Big] = \frac{d}{d\,\tau}\Big[ \bar C (\tau)\,e(\tau) \Big],\\
\]
\[
\frac{d}{d\,\tau}\,\Big[ \bar F^{(ab)}(\tau, \theta)\, K (\tau, \theta) \Big] = \frac{d}{d\,\tau}\Big[ \bar C (\tau)\,\chi(\tau) \Big].  \eqno (A.3)
\] 
Substitutions from (A.2) and (62) lead to the following equations in terms of the fermionic
(anti-)ghost variables as well as  secondary variables $\bar f_{2}$ and  $\bar b_1$, namely;
\[
\dot {\bar C}\,\bar f_2 + \bar C\, \dot {\bar f}_2 - \bar C\, \ddot{\bar C}\,e - \bar C\, \dot {\bar C}\,\dot e = 0,
\]
\[
\dot {\bar C}\,\bar b_1 + \bar C\, \dot {\bar b}_1 - \bar C\, \ddot{\bar C}\,\chi - \bar C\, \dot {\bar C}\,\dot \chi = 0  \eqno (A.4)
\]
It is straightforward to verify that we have the following solutions: 
\[\bar f_2 = \bar C\,\dot e + \dot {\bar C}\, e, \qquad \bar b_1 = \bar C\,\dot \chi + \dot {\bar C}\, \chi. \eqno (A.5)
\]
We take now the anti-BRST invariance: $s_{ab}\,(\dot B\, \bar C - B\, \dot {\bar C}) = 0$. This leads to the following 
restrictions on the {\it chiral} supervariables
\[
 \dot {\tilde B}(\tau, \theta)\, \bar F^{(ab)} (\tau, \theta) -  {\tilde B}(\tau, \theta)\, \dot {\bar F}^{(ab)} (\tau, \theta) = 
 \dot B (\tau )\,\bar C (\tau ) -    B (\tau )\, \dot {\bar C} (\tau ). \eqno (A.6) 
\]
Substitutions of the super expansions from (62) and (A.2) lead to the following relationship
\[
\dot B\, \bar C\,\dot {\bar C} - B\, \bar C \, \ddot {\bar C} + \dot {\bar {f_1}}\, \bar C - \bar f_1\,\dot {\bar C} = 0.     \eqno (A.7)
\]
It is evident that the precise solution is $\bar {f_1} = \dot B\, \bar C - B\, \dot {\bar C}$. Thus far, we have been able to determine
the precise forms of the secondary variables ($\bar b_1, \bar b_3, \bar f_1, \bar f_3$) in terms of the basic, auxiliary and (anti-)ghost
variables of the Lagrangians (24).

In our present Appendix, we have followed the tricks and techniques of ACSA to BRST formalism which was motivated by our standard
assumption that: $s_b \bar C = i\, B, \, s_{ab}\, C = i\, \bar B$ in the BRST approach to gauge and/or diffeomorphism invariant
theories. This standard assumption led to the (anti-)chiral super expansions of the supervariables in Eq. (47). This implies
that we have already determined the remaining secondary variable of Eq. (62) as: $\bar b_2 = i\, \bar B$. This completes our
discussion on the step-by-step determination of the secondary variables of [cf. Eq. (63)] which are present in the chiral
super expansions (62).\\

\begin{center}
{\bf Appendix B: On the Proof of Eqs. (28) and (74) in the Ordinary Space}\\
\end{center}
\vskip 0.5cm
For the sake of completeness, we provide here the explicit proof of the Eqs. (28) and/or (74) in the {\it ordinary} space that leads to 
the derivation of the CF-type restriction: $B + \bar B + i\, (\bar C \,\dot C - \dot {\bar C}\, C) = 0$ which demonstrates, thereby,
the equivalence of the Lagrangians $L_B$ and $L_{\bar B}$ w.r.t. the (anti-)BRST symmetry transformations. This is due to our earlier 
observation  [cf. Eq. (27)] that $L_{\bar B}$ has a {\it perfect} symmetry w.r.t. the anti-BRST symmetry transformations $s_{ab}$.
The direct applications of $s_b$ on $L_{\bar B}$ leads to the following: 
\[
s_b L_{\bar B}  = \frac {d}{d\tau}\Big[C\,L_{f}  - e^2\,(\bar B\, \dot C + i\,\dot{\bar C}\,\dot C\, C) - e\,\dot e \,(\bar B\, C + i\,{\bar C}\,\dot C\,C) \Big]
+ e^2\,[\dot {\bar B}\,\dot C + \dot B\,\dot C + i\,\bar C \ddot C\,\dot C - i\,\ddot{\bar C}\,C\,\dot C]\]
\[+ e\,\dot e\,[B\,\dot C + \bar B\,\dot C - i\,\dot {\bar C}\,C\,\dot C] + \bar B\,\dot{\bar B}\, C - \bar B\,\dot B\, C 
- \bar B ^2\,\dot C - 2\,B\,\bar B\,\dot C
-i\,\dot B\,\bar C\dot C\, C + i\,B\,\dot {\bar C}\,\dot C\, C \]
\[+ 2\,i\,\dot {\bar B}\,\bar C\, \dot C\, C + 2\,i\,\bar B\, \bar C\, \ddot C\, C + 2\,i\,\bar B\,\dot{\bar C}\,\dot C\,C.
 \eqno (B.1)\]
It is straightforward to note (with the input ${\dot C}^2 = 0$) that the coefficients of $e^2$ and $e\,\dot e$ can be expressed in terms of the CF-type restrictions 
$[B + \bar B + i\,(\bar C\,\dot C - \dot{\bar C}\,C) = 0]$ as follows: 
\[
e^2\Big[\frac {d}{d\tau}\Big\{B + \bar B + i\,(\bar C\,\dot C - \dot{\bar C}\,C)\Big\} \Big]\,\dot C
 + e\,\dot e\,\Big[B + \bar B + i\,(\bar C\,\dot C - \dot{\bar C}\,C)\Big]\, \dot C.
 \eqno (B.2)\]
 At this stage, we focus on the terms $\bar B\,\dot {\bar B}\, C - \bar B\, \dot B\,C - \bar B^2 \, \dot C 
 - 2\,B\, \bar B\,\dot C $ [from (B.1)] which can be expressed as a sum of a total derivative and other terms, namely;
\[
\frac {d}{d\tau} \Big[\bar B^2\, C\Big]  - (\dot B + \dot {\bar B})\,\bar B\, C - 2\,\bar B^2\, \dot C  - 2\,B\,\bar B\, \dot C
 = \frac {d}{d\tau} \Big[\bar B^2\, C\Big] - [\dot B + \dot {\bar B} + i\, (\bar C\, \ddot C - \ddot {\bar C}\, C]\, \bar B\, C\]
\[ + \,i\, \bar B\, \bar C\ddot C\, C - 2\,\bar B\,\dot C\, [B  + \bar B  + i\, (\bar C\, \dot C  - \dot {\bar C}\, C)] + 2\, i\, \bar B\, \dot {\bar C}\, \dot C\, C. 
\eqno (B.3)\]
Now adding the left-over terms {\it without} the total derivative terms as well as the CF-type terms from (B.1) and (B.3), we obtain the following:
\[
 2\, i\, \bar B\, \bar C \, \ddot C \, C  -  i\, \dot B \, \bar C\,\dot C \, C +  i\,  B \, \dot {\bar C}\,\dot C \, C  + 2\, i\, \bar B \, \dot {\bar C}
 \, \dot C \, C +  i\, \bar B \, \bar C\,\ddot C \, C\]
\[ + 2\, i\, \bar B\, \dot{\bar C} \, \dot C \, C + 2\,i\,\dot {\bar B}\,\bar C\, \dot C\, C. 
\eqno (B.4)\]
It can be readily seen that the following is 
completely true if we express the {\it first} two terms $(2\, i\, \bar B\, \bar C\, \ddot C \, C - i\, \dot B \,\bar C \,\dot C \, C)$
of the above equation, namely;
\[
\frac {d}{d\tau} \Big [ 2\, i\, \bar B \bar C \, \dot C \, C  -  i\, B \, \bar C \dot C \, C \Big] -  2\, i\, \bar B \dot {\bar C }\, \dot C\, C
-2\,i\,\dot {\bar B} \, \bar C \, \dot C \, C  + i\, B\, \dot {\bar C}\, \dot C \, C + i\, B \, \bar C \,\ddot C \, C.  
\eqno (B.5)\]
At this juncture, we add the terms from (B.4) and non-derivative terms from (B.5) to yield the following result
\[
2\,i\, (B + \bar B)\, \dot {\bar C}\, \dot C\, C + i\,(B + \bar B)\, \bar C\ddot C \, C,
\eqno (B.6)\]
which can be re-expressed, in terms of the CF-type restriction, as follows:
\[
2 \, i\,[B  + \bar B  + i\, (\bar C\,\dot C - \dot {\bar C}\, C)]\, \dot {\bar C}\, \dot C \, C 
+ i\, [B  + \bar B  + i\, (\bar C\,\dot C - \dot {\bar C}\, C)]\, \bar C\, \ddot C\, C.
\eqno (B.7)\] 
The total sum of the contributions from  (B.3), (B.5) and (B.7) is equal to: 
\[
\frac {d}{d\tau}\Big [ i \, (2\, \bar B - B)\, \bar C\, \dot C \, C + \bar B^2\, C \Big] - \frac {d}{d\tau}\Big [B  + \bar B  
+ i\, (\bar C\,\dot C - \dot {\bar C}\, C)\Big]\, \bar B \, C \]
 \[  + \big [B  + \bar B  + i\, (\bar C\,\dot C - \dot {\bar C}\, C)\big]\, 
\big [2\, i\, \dot {\bar C}\, \dot C\, C + i\, \bar C\,\ddot C\, C - 2\, \bar B \, \dot C \big].
\eqno (B.8)\]
Now adding all the terms of equations (B.1), (B.2) and (B.8), we obtain the {\it same} result as given in Eqs. (28) and/or (74) in the 
{\it ordinary} space for the computation of $s_b\, L_{\bar B}$.

We wrap-up this Appendix with the following concluding remarks. We have already demonstrated the existence of the CF-type restriction in the 
{\it ordinary space} [cf. Eq. (28)] and {\it superspace} [cf. Eq. (74)] by  expressing the BRST symmetry transformation (i.e. $s_b L_{\bar B}$)
of $L_{\bar B}$. Exactly in a similar manner, it can be shown that the quantity $ s_{ab} L_B$ can be
expressed in the {\it ordinary} space [cf. Eq. (29)] and superspace [cf. Eq. (72)]
leading to the appearance of the CF-type restriction. It is interesting, furthermore,  to point out that {\it this} restriction also appears when we prove 
the absolute anticommutativity of the conserved and off-shell nilpotent 
(anti-)BRST charges in the {\it ordinary} space as well as in the {\it superspace} (cf. Appendix D for details) using the ACSA to BRST formalism.\\

\begin{center}
{\bf Appendix C: On the Different Forms of the Conserved (Anti-)BRST Charges}\\
\end{center}
\vskip 0.5cm
Besides the expressions for the (anti-)BRST charges in (31) and (32), we require different forms of these charges to prove 
their nilpotency (i.e. 
fermionic nature) and anticommutativity (i.e. linear independence) in a straightforward manner. This exercise has been found to be 
 advantageous in the
context of our discussion on ACSA, too. First and foremost, we concentrate on the derivation of the BRST charge in (77) and its 
usefulness. In this connection, we note that the Noether charge (31) can be re-written as follows
\[
Q_B \rightarrow  \tilde Q_B = e^2\,[B\,\dot C - \dot B\,C - i\,\dot{\bar C}\,\dot C \, C] + B^2\,C + e \, \dot e \, B\,C
 - i\, B \,\bar C\,\dot C \, C,    \eqno (C.1)
\] 
where we have used the following EL-EOMs (and their modified version), namely;
\[
p_\mu \, \psi^\mu = m\, \psi_5, \qquad \quad  \frac {e\,C}{2}\,(p^2 - m^2) = -e^2\,\dot B\,C - i\,e^2\,\dot{\bar C}\,\dot C\,C.
   \eqno (C.2)
\]
At this juncture, we note that $B = - e\,\dot e + i\,(2\, \dot {\bar C}\, C + \bar C \, \dot C)$.
The substitution of this expression leads to the following observations, namely;
\[
B^2\,C + e\,\dot e \,B\,C = -\,i\,e\,\dot e \,\bar C\,\dot C\, C, \qquad \quad -\,i\,B\,\bar C\, \dot C\, C = i\,e\,\dot e \,\bar C\,\dot C\, C.   \eqno (C.3)
\]
Thus, we obtain the modified form of the BRST charge $(\tilde Q_B)$ as 
\[
\tilde Q_B \rightarrow \tilde Q_{B}^{(1)} = e^2\,[B\,\dot C - \dot B\,C - i\,\dot{\bar C}\,\dot C \, C].    \eqno (C.4)
\]
This is because of the fact that the sum of last {\it three} terms in (C.1) is zero. It can be readily checked that the above form of the charge [cf. Eq. (C.4)] is 
{\it not} off-shell nilpotent. Hence, it is {\it not} suitable for our further discussions.

Let us focus on the derivation of an alternative form of this BRST charge. In the expression for the charge $\tilde Q_B$ [cf. Eq. (C.1)], 
we can express $[-\,i\,B\,\bar C\,\dot C\,C]$, from the {\it third} equation from the top in (76) which yields the following
\[
-\,i\,B\,\bar C\,\dot C\,C = i\, e^2\,\bar C\,\ddot C\, C + 3\,i\,e\,\dot e\,\bar C\,\dot C\,C.   \eqno (C.5)
\]
Using (C.3) and (C.5), we obtain
\[
{\tilde Q}_B^{(1)} \quad \longrightarrow \quad \tilde Q_{B}^{(2)} = e^2\,[B\,\dot C - \dot B\, C + i\,(\bar C \,\ddot C\, C 
- \dot{\bar C}\,\dot C \, C)] + 2\,i\,e\,\dot e\, \bar C\, \dot C\, C.  \eqno (C.6)
\]
It is elementary exercise to check that the above expression for the BRST charge is {\it still} not off-shell nilpotent of order two. Hence, 
it can{\it not} be expressed as an exact quantity w.r.t. the BRST transformations $(s_b)$. At this crucial point, we re-express (C.5) in a
different type of mathematical form as follows  
\[
i\,B\,\bar C\,\dot C\,C = - i\, e^2\,\bar C\,\ddot C\, C -  2\,i\,e\,\dot e\,\bar C\,\dot C\,C - i\,e\,\dot e\,\bar C\,\dot C\,C.   \eqno (C.7)
\]
From the relationship: $B = -\,e\,\dot e + i\,(2\,\dot{\bar C}\,C + \bar C\,\dot C)$, it is clear that $i\,\bar C\,\dot C = 
B + e\,\dot e - 2\,i\,\dot{\bar C}\,C$. As a result, we have the following
\[
i\,B\,\bar C\,\dot C\,C = B\,(B + e\,\dot e  - 2\,i\,\dot{\bar C}\,C)\,C \quad \equiv \quad B^2\,C + e\,\dot e\,B\,C.   \eqno (C.8)
\]
Substituting the above equality into (C.7), we obtain 
\[
B^2\,C + e\,\dot e\,B\,C + i\,e\,\dot e\,\bar C\,\dot C\,C = -\,i\,e^2\,\bar C\,\ddot C\,C - 2\,i\,e\,\dot e\,\bar C\,\dot C\,C. 
 \eqno(C.9)
\]
However, we also note that $i\,e\,\dot e\,\bar C\,\dot C\,C = -\,i\,B\,\bar C\,\dot C\,C$ [cf. Eq. (C.3)] due to the fact that 
$B = -e\,\dot e + i\,(2\,\dot{\bar C}\,C + \bar C\,\dot C)$. Thus, we observe that the equality in  (C.9) reduces to: 
\[
B^2\,C + e\,\dot e\,B\,C - i\,B\,\dot{\bar C}\,\dot C\,C = -\,i\,e^2\,\bar C\,\ddot C\,C - 2\,i\,e\,\dot e\,\bar C\,\dot C\,C. 
 \eqno(C.10)
\]
The above equation leads to the following
\[
{\tilde Q}_B  \rightarrow  Q_B^{(1)} = e^2\,\Big[B\,\dot C - \dot B\,C - i\,(\dot{\bar C}\,\dot C\,C + \bar C\,\ddot C\,C)\Big]
- 2\,i\,e\,\dot e\,\bar C\,\dot C\,C, \eqno(C.11)
\]
where ${\tilde Q}_B$ is quoted in (C.1) and $Q_B^{(1)}$ is written in Eq. (77) of the main body of our text. The importance of (C.11)
is the observation that it can be written as an exact quantity w.r.t. the BRST transformations $(s_b)$. As a result, the
charge $Q_B^{(1)}$ is off-shell nilpotent of order two (i.e. $s_b\,Q_B^{(1)} = -\,i\,\{Q_B^{(1)}, Q_B^{(1)}\} = 0 \Longrightarrow [Q_B^{(1)}]^2 = 0)$. 
Exactly similar kinds of arguments can be provided for the derivation of $Q_{\bar B}^{(1)}$ in Eq. (77). As a consequence, we observe that: 
$s_{ab}\,Q_{\bar B}^{(1)} =
 -\,i\,\{Q_{\bar B}^{(1)}, Q_{\bar B}^{(1)}\} = 0 \Longrightarrow [Q_{\bar B}^{(1)}]^2 = 0$. This observation proves the off-shell nilpotency of 
$Q_{\bar B}^{(1)}$ in a straightforward fashion.  \\

\begin{center}
{\bf Appendix D: On an Alternative Proof of the Existence of the CF-Type Restriction on Our SUSY System}\\
\end{center}
\vskip 0.5cm
Our present Appendix provides an alternative proof of the existence of the CF-type restriction: $B + \bar B + i\,(\bar C\,\dot C - \dot{\bar C}\,C) = 0$
which is straightforward {\it and} different from our derivation in Sec. 7 where the proof is a bit {\it subtle}. In this context, we apply, first of all,
directly the BRST symmetry transformations $(s_b)$ on the expression for the anti-BRST charge $Q_{\bar B}^{(1)}$ [cf. Eq. (77)]. In other words, 
we derive $s_b\,Q_{\bar B}^{(1)} = -\,i\,\{Q_{\bar B}^{(1)}, Q_{B}^{(1)}\}$ which is nothing but the anticommutator of the anti-BRST
charge $(Q_{\bar B}^{(1)})$ {\it with} the BRST charge $(Q_B^{(1)})$. The outcome of this exercise can be explicitly expressed as:
\[
s_b\,Q_{\bar B}^{(1)} = 2\,e^2\,\bar B\,\dot{\bar C}\,C - 2\,e^2\,\dot{\bar B}\,\bar C\,\dot C - 2\,e\,\dot e\,\dot{\bar B}\,\bar C\,C + 2\,e\,\dot e\,\bar B
\,\dot{\bar C}\,C - e^2\,\ddot{\bar B}\,\bar C\,C + e^2\,\bar B\,\bar C\,\ddot C 
\]
\[
+ 2\,i\,e\,\dot e\,\dot{\bar C}\,\bar C\,\dot C\,C + e^2\,B\,\ddot{\bar C}\,C
- e^2\,\ddot B\,\bar C\,C +2\,e\,\dot e\,B\,\dot{\bar C}\,C - 2\,e\,\dot e\,\dot B\,\bar C\,C + i\,e^2\,\ddot{\bar C}\,\bar C\,\dot C\,C 
\]
\[
+ i\,e^2\,\dot{\bar C}
\bar C\,\ddot C\,C + e^2\,B\,\dot{\bar C}\,\dot C - e^2\,\dot B\,\bar C\,\dot C + i\,e^2\,B\,\dot{\bar B} - i\,e^2\,\bar B\,\dot B + e^2\,\dot{\bar B}\,\dot{\bar
 C}\,C - e^2\,\bar B\,\dot{\bar C}\,\dot C.\eqno(D.1)
\]
The above expression can be re-arranged in such a manner that we shall have the coefficients of $e^2$ and $2\,e\,\dot e$ separately and independently as
illustrated below:
\[
s_b\,Q_{\bar B}^{(1)} = 2\,e\,\dot e\,\big[B\,\dot{\bar C}\,C + \bar B\,\dot{\bar C}\,C - \dot B\,\bar C\,C - \dot{\bar B}\,\bar C\,C - i\,\dot{\bar C}\,
\bar C\,\dot C\,C\big]
\] 
\[
+ e^2\,\Big[2\,(\bar B\,\dot{\bar C} - \dot{\bar B}\,\bar C)\,\dot C - i\,(\ddot{\bar C}\,\dot C + \dot{\bar C}\,\ddot C)\,\bar C\,C + (\ddot{\bar B}\,C
- \bar B\,\ddot C)\,\bar C - (\dot{\bar B}\,C - \bar B\,\dot C)\,\dot{\bar C}
\]
\[
 + i\,B\,\dot{\bar B} - i\,\bar B\,\dot B + (B\,\ddot{\bar C} - \ddot B\,\bar C)\,C + (B\,\dot{\bar C} - \dot B\,\bar C)\,\dot C\Big]. \eqno {(D.2)}
\]
At this stage, the straightforward algebraic exercise produces the following result
\[
s_b\,Q_{\bar B}^{(1)} = -\,i\,\{Q_{\bar B}^{(1)}, Q_B^{(1)}\}
\] 
\[
\equiv 2\,e\,\dot e\,\Big[\{B + \bar B + i\,(\bar C\,\dot C - \dot{\bar C}\,C)\}\,\dot{\bar C}\,C - \frac{d}{d\,\tau}\,\{B + \bar B + i\,(\bar C\,\dot C 
- \dot{\bar C}\,C)\}\,\bar C\,C\Big]
\]
\[
+ e^2\,\Big[\frac{d}{d\,\tau}\Big\{\big[B + \bar B + i(\bar C\,\dot C - \dot {\bar C}\,C)\big]\,(\dot{\bar C}\,C - i\,\bar B) 
-  \frac{d}{d\,\tau}\big[B + \bar B + i(\bar C\,\dot C - \dot {\bar C}\,C )\big]\bar C\,C\Big\}
\]
\[
+ 2\,i\,\dot{\bar B}\,\big[B + \bar B + i(\bar C\,\dot C - \dot {\bar C}\,C )\big]\Big], \eqno{(D.3)}
\]
which demonstrates {\it explicitly} that the absolute anticommutativity property $(\{Q_{\bar B}^{(1)}, Q_B^{(1)}\} = 0)$ is satisfied if and only
if the CF-type restriction: $B + \bar B + i\,(\bar C\,\dot C - \dot{\bar C}\,C) = 0$ is invoked from {\it outside}. In other words, the {\it requirement} 
of the absolute anticommutativity of the (anti-)BRST charges leads to the existence of a CF-type restriction on our theory.

We now concentrate on the computation  of $s_{ab}\,Q_{B}^{(1)} = -\,i\,\{Q_{ B}^{(1)}, Q_{\bar B}^{(1)}\}$ which leads to the following as the sum of 
the coefficients of $e^2$ and $2\,e\,\dot e$, namely;
\[
s_{ab}\,Q_B^{(1)} = 2\,e\,\dot e\,\Big[B\,\bar C\,\dot C - \dot B\,\bar C\,C + \bar B\,\bar C\,\dot C - \dot{\bar B}\,\bar C\,C - i\,\dot{\bar C}\,\bar C
\,\dot C\,C\Big] 
\]
\[
+ e^2\,\Big[2\,B\,\dot{\bar C}\,\dot C - 2\,\dot B\,\dot{\bar C}\,C - i\,\dot{\bar C}\,\bar C\,\ddot C\,C - i\,\ddot{\bar C}\,\bar C\,\dot C\,C - \ddot B\,\bar C
\,C + B\,\ddot{\bar C}\,C + \dot B\,\bar C\,\dot C - B\,\dot{\bar C}\,\dot C \]
\[
- i\,\dot B\,\bar B + i\,B\,\dot{\bar B} + \bar B\,\dot{\bar C}\,\dot C - \dot{\bar
B}\,\dot{\bar C}\,C + \bar B\,\bar C\,\ddot C - \ddot{\bar B}\,\bar C\,C\Big]. \eqno(D.4)
\]
The above expression can be re-arranged by performing some algebraic tricks in the following form (in terms of the CF-type 
restrictions  $B + \bar B + i\,(\bar C\,\dot C - \dot{\bar C}\,C) = 0$), namely;
\[
s_{ab}\,Q_{B}^{(1)} = e^2\Big [\frac{d}{d\,\tau}\Big\{\big[B + \bar B + i(\bar C\,\dot C - \dot {\bar C}\,C)\big]\,(i\,B + \bar C\,\dot C) 
-  \frac{d}{d\,\tau}\big[B + \bar B + i(\bar C\,\dot C - \dot {\bar C}\,C )\big]\bar C\,C\Big\} 
\]
\[
- 2\,i\,\dot B\,\big[B + \bar B + i(\bar C\,\dot C - \dot {\bar C}\,C )\big]\Big] + 2\,e\,\dot e\,\bar C\Big[\big(B + \bar B + 
i(\bar C\,\dot C - \dot {\bar C}\,C)\big)\,\dot C
\]
\[
-\frac{d}{d\,\tau}\,\big[B + \bar B + i(\bar C\,\dot C - \dot {\bar C}\,C)\big]\,C\Big].     \eqno (D.5)
\]
The above expression demonstrates that the absolute anticommutativity property (i.e. $s_{ab}\,Q_B^{(1)} = -\,i\,\{Q_B^{(1)}, Q_{\bar B}^{(1)}\} = 0$)
is satisfied if and only if the validity of the CF-type restriction is invoked at the {\it quantum} level in our BRST {\it quantized} theory. There is 
another way of stating this observation. That is to say the fact that the sanctity of anticommutativity property (i.e. $\{Q_B^{(1)}, Q_{\bar B}^{(1)}\} = 0$)
 leads to the deduction of the CF-type restriction in a straightforward manner.

We close our Appendix with the final remark that we can capture the derivation of the CF-type restriction within the purview of ACSA.
Towards this central objective in our mind, first of all, we generalize the expressions for the BRST and anti-BRST charges [cf. Eq. (77)] on the $(1, 1)$-
dimensional {\it chiral} and {\it anti-chiral} supermanifold as  
\[
Q_B^{(1)} \longrightarrow {\tilde Q}_B^{(1c)}(\tau, \theta) = E^{(ab)}(\tau, \theta)\,E^{(ab)}(\tau, \theta)\,\Big[{\tilde B}^{(ab)}(\tau, \theta)
\,{\dot F}^{(ab)}(\tau, \theta) - \dot{\tilde B}^{(ab)}(\tau, \theta)\,F^{(ab)}(\tau, \theta) 
\]
\[
- i\,\big({\bar F}^{(ab)}(\tau, \theta)\,{\ddot F}^{(ab)}(\tau, \theta)\,F^{(ab)}(\tau, \theta) + \dot{\bar F}^{(ab)}(\tau, \theta)\,{\dot F}^{(ab)}
(\tau, \theta)\,F^{(ab)}(\tau, \theta)\big)\Big]
\]
\[
-\,2\,i\,E^{(ab)}(\tau, \theta)\,\dot E^{(ab)}(\tau, \theta)\,{\bar F}^{(ab)}(\tau, \theta)\,{\dot F}^{(ab)}
(\tau, \theta)\,F^{(ab)}(\tau, \theta),
\]

\[
Q_{\bar B}^{(1)} \longrightarrow {\tilde Q}_{\bar B}^{(1ac)}(\tau, \bar\theta) = E^{(b)}(\tau, \bar\theta)\,E^{(b)}(\tau, \bar\theta)\,
\Big[\dot{\tilde{\bar  B}}^{(b)}(\tau, \bar\theta)
\,{\bar F}^{(b)}(\tau, \bar\theta) - \tilde{\bar B}^{(b)}(\tau, \bar\theta)\,\dot{\bar F}^{(b)}(\tau, \bar\theta) 
\]
\[
+ \,i\,\big(\ddot{\bar F}^{(b)}(\tau, \bar\theta)\,{\bar F}^{(b)}(\tau, \bar\theta)\,F^{(b)}(\tau, \bar\theta) + \dot{\bar F}^{(b)}(\tau, \bar\theta)\,{\bar
 F}^{(b)}(\tau, \bar\theta)\,{\dot F}^{(b)}(\tau, \bar\theta)\big)\Big]
\]
\[
+\, 2\,i\,E^{(b)}(\tau, \bar\theta)\,\dot E^{(b)}(\tau, \bar\theta)\,\dot{\bar F}^{(b)}(\tau, \bar\theta)\,{\bar F}^{(b)}
(\tau, \bar\theta)\,F^{(b)}(\tau, \bar\theta),  \eqno (D.6)
\]
where the superscripts $(1c)$ and $(1ac)$, on the BRST and anti-BRST charges, denote the {\it chiral} and {\it anti-chiral} versions of Eq. (77).
The other notations have already  been explained earlier. It can now be checked that we have the following:
\[
\frac{\partial}{\partial\,\theta}\,{\tilde Q}_B^{(1c)}(\tau, \theta) = e^2\Big [\frac{d}{d\,\tau}\Big\{\big[B + \bar B + i(\bar C\,\dot C - \dot {\bar C}\,C)\big]\,(i\,B + \bar C\,\dot C) 
-  \frac{d}{d\,\tau}\big[B + \bar B + i(\bar C\,\dot C - \dot {\bar C}\,C )\big]\bar C\,C\Big\} 
\]
\[
- 2\,i\,\dot B\,\big[B + \bar B + i(\bar C\,\dot C - \dot {\bar C}\,C )\big]\Big] + 2\,e\,\dot e\,\bar C\Big[\big(B + \bar B + 
i(\bar C\,\dot C - \dot {\bar C}\,C)\big)\,\dot C
\]
\[
-\frac{d}{d\,\tau}\,\big[B + \bar B + i(\bar C\,\dot C - \dot {\bar C}\,C)\big]\,C\Big] \Longleftrightarrow   s_{ab}\,Q_B^{(1)} = -\,i\,\{Q_B^{(1)}, Q_{\bar B}^{(1)}\},      
\]

\[
\frac{\partial}{\partial\,\bar\theta}\,{\tilde Q}_{\bar B}^{(1ac)}(\tau, \bar\theta) = 2\,e\,\dot e\,\Big[\{B + \bar B + i\,(\bar C\,\dot C - \dot{\bar C}\,C)\}\,\dot{\bar C}\,C - \frac{d}{d\,\tau}\,\{B + \bar B + i\,(\bar C\,\dot C 
- \dot{\bar C}\,C)\}\,\bar C\,C\Big]
\]
\[
+ e^2\,\Big[\frac{d}{d\,\tau}\Big\{\big[B + \bar B + i(\bar C\,\dot C - \dot {\bar C}\,C)\big]\,(\dot{\bar C}\,C - i\,\bar B) 
-  \frac{d}{d\,\tau}\big[B + \bar B + i(\bar C\,\dot C - \dot {\bar C}\,C )\big]\bar C\,C\Big\}
\]
\[
+ 2\,i\,\dot{\bar B}\,\big[B + \bar B + i(\bar C\,\dot C - \dot {\bar C}\,C )\big]\Big] \Longleftrightarrow  s_b\,Q_{\bar B}^{(1)} = -\,i\,\{Q_{\bar B}^{(1)}, Q_{B}^{(1)}\}.
 \eqno(D.7)
\]
Thus, a careful and close look at the r.h.s. of (D.7) demonstrates that we have 
deduced the CF-type restriction: $B + \bar B + i\,(\bar C\,\dot C 
- \dot {\bar C}\,C) = 0$ within the ambit of ACSA in the proof of the absolute 
anticommutativity of the conserved (anti-)BRST charges. It is worthwhile
to mention here that our equation (D.7) captures the absolute anticommutativity 
of the (anti-)BRST charges $[Q_{(\bar B)B}]$ in the {\it ordinary} space
as well as in the {\it superspace} provided the whole theory is considered on the subspace 
(of the entire quantum Hilbert space) of variables where the CF-type restriction:
$B + \bar B + i\,(\bar C\,\dot C - \dot {\bar C}\,C) = 0$ is satisfied.

\vskip 1.3cm

\noindent
{\bf Acknowledgments}\\

\noindent
B. Chauhan thankfully acknowledges the financial support from DST (Govt. of India) under its
INSPIRE-fellowship scheme {\it and}  A. Tripathi as well as A. K. Rao are grateful 
to the general BHU-fellowship [from Banaras Hindu University (BHU), Varanasi] under which
the present investigation has been carried out.\\

\noindent
{\bf\large Data Availability}\\

\noindent
No data were used to support this study. \\

\vskip 0.4 cm

\noindent
{\bf\large Conflicts of Interest}\\

\noindent
The authors declare that there is no conflicts of interest.\\

\end{document}